\documentclass[a4paper,11pt]{article}
\usepackage{jheppub} 

\newmuskip\pFqmuskip
\newcommand*\pFq[6][8]{%
  \begingroup 
  \pFqmuskip=#1mu\relax
  \mathchardef\normalcomma=\mathcode`,
  \mathcode`\,=\string"8000
  \begingroup\lccode`\~=`\,
  \lowercase{\endgroup\let~}\pFqcomma
F{\left(\genfrac..{0pt}{}{#4}{#5};#6\right)}%
  \endgroup
}
\newcommand{\pFqcomma}{{\normalcomma}\mskip\pFqmuskip}

\newcommand*\pFfulq[6][8]{%
  \begingroup 
  \pFqmuskip=#1mu\relax
  \mathchardef\normalcomma=\mathcode`,
  \mathcode`\,=\string"8000
  \begingroup\lccode`\~=`\,
  \lowercase{\endgroup\let~}\pFqcomma
{}_{#2}F_{#3}{\left(\genfrac..{0pt}{}{#4}{#5};#6\right)}%
  \endgroup
}

\arxivnumber{2504.03435} 

\title{\boldmath Exactly solvable models
for universal operator growth}

\author[a,1]{Oleksandr Gamayun\note{Corresponding author.}}
\affiliation[a]{London Institute for Mathematical Sciences\\Royal Institution, 21 Albemarle St, London W1S 4BS, UK}
\author[b,2]{Murtaza Ali Mir}
\affiliation[b]{Centre of Excellence, Chinar Quantum AI, \\
	F3 Uptown Tower, Hyderpora, Srinagar, India
}
\author[c,3]{Oleg Lychkovskiy}
\affiliation[c]{Castiglioncello, Italy}

\author[d,4]{Zoran Ristivojevic}
\affiliation[d]{Laboratoire de Physique Theorique, Universit de Toulouse, CNRS
\\118 Route de Narbonne, 31062 Toulouse Cedex 9, France}

\emailAdd{og@lims.ac.uk}
\emailAdd{murtaza.mir@skoltech.ru}
\emailAdd{lychkovskiy@gmail.com}
\emailAdd{zoran.ristivojevic@utoulouse.fr}

\abstract{Quantum observables of generic many-body systems exhibit a universal pattern  of growth in the Krylov space of operators. This pattern becomes particularly manifest in the Lanczos basis, where the evolution superoperator assumes the tridiagonal form. According to the universal operator growth hypothesis, the nonzero elements of the superoperator, known as Lanczos coefficients, grow asymptotically linearly.  We introduce and explore broad families of Lanczos coefficients that are consistent with the universal operator growth and lead to the exactly solvable dynamics. Within these families, the subleading terms of asymptotic expansion of the Lanczos sequence can be controlled and fine-tuned to produce diverse dynamical patterns. For one of the families, the  Krylov complexity is computed exactly.}

\begin{document}
\maketitle
\flushbottom

\section{Introduction: recursion method and universal operator growth}
\label{sec:intro}

Describing the dynamics of quantum many-body systems is among the main objectives of condensed matter and quantum field theories. This task has a reputation of being extremely complex {\it in general}. The most well-understood tractable scenario emerges if a system under study is close in some sense to a collection of noninteracting (quasi)particles. In this case a diverse and sophisticated toolbox of perturbative techniques is available and often sufficient for an exhaustive quantitative description. The opposite case of a generic quantum many-body system far from any free-particle point has been until recently widely believed to be intractable (apart from some exceptional types of systems and special techniques, including AdS/CFT correspondence \cite{Liu_2020_Holographic}, low-dimensional systems solvable by matrix product state methods \cite{Banuls_2009_Matrix,orus2019tensor}, etc.). This assessment is being revised nowadays, in large part thanks to the universal operator growth hypothesis proposed in ref.~\cite{Parker_2019}. The latter behavior emerges in the framework of the recursion method developed long ago \cite{Mori_1965_Continued-fraction,Dupuis_1967_Moment,viswanath2008recursion}.

In essence, the recursion method amounts to solving coupled Heisenberg equations of motion in the tridiagonal Lanczos basis in operator space. For our purposes, it is enough to briefly outline the outcome of  the recursion method without going into derivations and technical details \cite{viswanath2008recursion,Parker_2019,Muck_2022_Krylov}. We consider a many-body system described by a local Hamiltonian $H$ and focus on the most basic object within the method  -- an infinite-temperature autocorrelation function. It is defined as $C(t)= \mathrm{Tr}\,\{\mathcal{O}(t)\mathcal{O}(0)\}/\mathrm{Tr}\,\{\mathcal{O}^2\}$, where $\mathcal{O}$ is some local observable that does not depend on time in the Schr\"{o}dinger picture. In the Heisenberg picture, it depends on time and evolves according to the Heisenberg equation of motion, $\partial_t \mathcal{O}(t) =i[H,\mathcal{O}(t)]$. Here and in the following we set the reduced Planck constant to one, $\hbar=1$. In order to compute the autocorrelation function, it turns out that it is sufficient to solve the discrete Schr\"{o}dinger equation
\begin{align}\label{system}
    \partial_t \varphi_n(t)& =-b_{n+1}\,\varphi_{n+1}(t)+ b_n\,\varphi_{n-1}(t)\,,\qquad n=0,1,2,\dots\,,
\end{align}
subject to the conditions
\begin{align}\label{eq:bc}
\varphi_n(0)  =  \delta_{n,0}\,,\qquad \varphi_{-1}(t)=0\,.
\end{align}
Equation (\ref{system}) describes a fictitious single particle on a semi-infinite discrete interval, which is described by the wave function $\varphi_n(t)$. At $t=0$ it is in the origin and propagates in time. The set of positive coefficients $b_n \geq 0$ in eq.~(\ref{system}) are known as Lanczos coefficients and play a major role. The solution of eq.~(\ref{system}) is a set of real functions $\varphi_n(t)$ for $n=0,1,2,\ldots$. It satisfies the normalization condition $\sum_{n=0}^{\infty}\varphi_n(t)^2=1$. At $t=0$, this is obvious due to the initial condition (\ref{eq:bc}). At $t>0$, differentiating the normalization condition it follows $\sum_{n=0}^{\infty}\varphi_n(t)\partial_t\varphi_n(t)=0$, which is satisfied due eq.~(\ref{system}). Once eq.~(\ref{system}) is solved, the autocorrelation function is simply given by
\begin{equation}
C(t)=\varphi_0(t)\,.
\end{equation}
We note that $C(0)=1$.

Lanczos coefficients $b_n$ of eq.~(\ref{system}) depend on the Hamiltonian $H$ and the operator $\mathcal{O}$. They can be obtained by the orthogonalization of the sequence of operators $\mathcal{L}^n\mathcal{O}$ for $n=0,1,2,\ldots$, and represent the norms of the orthogonal set of operators. Here $\mathcal{L}$ denotes the Liouville superoperator that acts on operators by the commutator as $\mathcal{L}\mathcal{O}=[H,\mathcal{O}]$. Therefore, in order to obtain  $\mathcal{L}^n\mathcal{O}$ for a given $n$, nested commutators should be evaluated. The computational complexity of this task grows exponentially with $n$, and therefore, in practice, only a finite number $n_{\rm{max}}$  of Lanczos coefficients is typically available.  As a consequence, eq.~(\ref{system})  cannot be solved. While the truncation of this equation at $n =n_{\rm{max}}$ provides an accurate approximation for $C(t)$ at short times, it breaks down at longer times. This was a severe limiting factor of the recursion method for decades \cite{viswanath2008recursion}.

The crucial step to resolve the above stalemate was made in ref.~\cite{Parker_2019} (see also a precursor work \cite{Liu_1990_Infinite-temperature,Florencio_1992_Quantum,Zobov_2006_Second,Elsayed_2014_Signatures,Bouch_2015_Complex}), where
a universal operator growth hypothesis has been put forward. It states that for a generic (in particular, nonintegrable)  many-body system,
Lanczos coefficients grow asymptotically linearly with $n$ (with an additional logarithmic correction in one-dimensional systems -- a case not addressed  in the present paper):
\begin{equation}\label{UOGH}
b_n=\alpha\;\! n +{o}(n) \,,\qquad n\rightarrow \infty \,.
\end{equation}
The universal operator growth hypothesis (\ref{UOGH}) has been subsequently confirmed \textit{explicitly} for various many-body models \cite{Cao_2021_Statistical,Noh_2021,Heveling_2022_Numerically,Uskov_Lychkovskiy_2024_Quantum,De_2024_Stochastic,Teretenkov_2025,Loizeau_2025_Quantum,Loizeau_2025_Codebase, shirokov2025quench,ermakov2025operator}. We note that the asymptotic  behavior (\ref{UOGH}) is typically slowed down for integrable systems \cite{Parker_2019}. The universal operator growth hypothesis implies the exponential growth of the {\it Krylov complexity} \cite{Parker_2019}
\begin{align}
\mathcal{K}(t)=\sum_{n=0}^{\infty}n\;\!\varphi_n(t)^2\,,
\end{align}
which is regarded to be a measure for the operator growth {\color{red} \cite{nandy2024quantum,baiguera2025quantum}}. The latter is an important and ubiquitous phenomenon that  appears in a variety of contexts ranging from quantum optics \cite{Caputa_2022_Geometry} and quantum networks \cite{Kim_2022_Operator} to cosmology \cite{Adhikari_2022_Cosmological}, black hole physics \cite{Jian_2021_Complexity,Kar_2022_Random}, holography \cite{Adhikari_2023_Krylov,Avdoshkin_2024_Krylov,chapman_krylov_2025,caputa2024krylov},  quantum billiards \cite{camargo2024spectral}  and conformal field theories \cite{Dymarsky_2021_Krylov,Caputa_2021_Operator,Caputa_2022_Geometry,Kar_2022_Random,Adhikari_2023_Krylov}.

The universal operator growth hypothesis \eqref{UOGH} signals that the recursion method might be augmented by replacing the unknown  Lanczos coefficients $b_n$ for $n> n_{\rm max}$, by their extrapolated counterparts $b_n^{\rm extr}=\alpha n$, where $\alpha$ is found by fitting the known Lanczos coefficients. This procedure can admit a perturbative guise as follows. One first introduces an  unperturbed Schr\"odinger equation of the form \eqref{system} with the coefficients exactly linear in $n$,
\begin{equation}\label{linear bn}
b_n^0=\alpha\;\!n \,, \qquad n=0,1,2,\ldots \,.
\end{equation}
The actual Schr\"odinger equation is obtained by perturbing each coefficient according to
$ b_n= b_n^0+\delta b_n$ with $\delta b_n=b_n-\alpha \;\!n$, cf., ref.~\cite{dodelson2025black}. The universal operator growth hypothesis \eqref{UOGH} guarantees that the perturbation $\delta b_n$  vanishes for large $n$. Note that a possibly large perturbation at a few first sites of the semi-infinite chain can be addressed separately \cite{Banchi_2013,Gamayun_2020,Ljubotina_2019,Parker_2019}. Importantly, the unperturbed dynamics governed by the Lanczos coefficients \eqref{linear bn} can be solved exactly \cite{Parker_2019}, leading to $C(t)=\mathrm{sech}(\alpha t)$.

While a perturbative scheme along these lines appears feasible, its actual implementation is far from being straightforward \cite{Parker_2019,Uskov_Lychkovskiy_2024_Quantum}. One particularly serious issue is that the subleading terms in the asymptotic expansion of $b_n$ (those hidden in $o(n)$ in eq.~\eqref{UOGH}) can have a strong impact on $\varphi_0(t)$, to the extent that the qualitative behavior of the autocorrelation function is altered \cite{Viswanath_1994_Ordering,Parker_2019,Yates_2020_Lifetime,Yates_2020_Dynamics,
Dymarsky_2021_Krylov,Bhattacharjee_2022_Krylov,Avdoshkin_2024_Krylov,Camargo_2023_Krylov,
Uskov_Lychkovskiy_2024_Quantum,dodelson2025black,lunt2025emergent,pinna2025approximation}.  It is therefore highly desirable to have a large set of sequences that satisfy the universal operator growth hypothesis \eqref{UOGH}, encompass various subleading terms, and lead to exactly solvable Schr\"odinger equation \eqref{system}. We will refer to such sequences satisfying the later requirement as {\it exactly solvable Lanczos sequences}. A wise choice of a suitable exactly solvable Lanczos sequence as a zero-order approximation could considerably improve the perturbative scheme for a particular many-body model.

The sequence \eqref{linear bn} is in fact a particular case of a more general exactly solvable sequence given by  \cite{Parker_2019}
\begin{equation}\label{bnsqrt}
b_n =\sqrt{n(n-1+\eta)} \,,
\end{equation}
with
\begin{equation}\label{mp}
    \varphi_n(t) = \sqrt{\frac{(\eta)_n}{n!}} \frac{\tanh^n(t)}{\cosh^\eta(t)}\,.
\end{equation}
Here $(\eta)_n = \Gamma(\eta+n)/\Gamma(\eta)$ is the Pochhammer symbol, and $\eta\ge1$ is a free parameter. By varying the parameter $\eta$, one can control the first subleading term in eq.~\eqref{bnsqrt} at large $n$. In eq.~(\ref{bnsqrt}) and in the following we have omitted the superscript from $b_n^0$ and set $\alpha=1$ for simplicity, which amounts to an appropriate rescaling of time. Indeed for the Lanczos coefficients $\tilde b_n=\alpha b_n$, the solution of eq.~(\ref{system}) is given by $\varphi_n(\alpha t)$, where $\varphi_n(t)$ is defined in eq.~(\ref{mp}). Many other examples of exactly solvable sequences are studied in ref.~\cite{Muck_2022_Krylov}, where some are in accordance with the universal operator growth hypothesis \eqref{UOGH} and some are not.

In the  present paper we introduce and analyze several  families of presumably unknown exactly solvable Lanczos sequences consistent with the universal operator growth hypothesis. We demonstrate that these multiparameter families enable fine tuning of the subleading terms and lead to autocorrelation functions with a diverse qualitative behavior. The remaining part of the paper is organized as follows. In section \ref{sec: recursion method} we provide a brief recap of the recursion method and its relation to the orthogonal polynomials. A family of models based on the continuous Hahn polynomials is introduced and explored in section \ref{sec: Hahn}. In section \ref{sec: alternating} we introduce another family of models whose distinctive feature is the sign alteration in the subleading terms. In section \ref{sec: nonzero equilibrium} we explore the Lanczos coefficients for the correlation functions with nonzero stationary late-time values. In section  \ref{sec: discussion} some broader  implications of the obtained  rigorous results  are discussed. Some technical details about the continuous Hahn polynomials are given in appendix \ref{appendix}.

\section{Recursion method and orthogonal polynomials \label{sec: recursion method}}

Equation~\eqref{system} can be solved by employing the method of orthogonal polynomials  \cite{Mori_1965_Continued-fraction,Dupuis_1967_Moment,Muck_2022_Krylov,sasaki2024towards}. In order to achieve that, we introduce the system of polynomials satisfying the standard recurrence relation of the form
\begin{equation}\label{fav}
	x \pi_n(x) = b_{n+1}\pi_{n+1}(x)+b_n\pi_{n-1}(x)\,,
\end{equation}
where $b_n$ is a given set of the coefficients. The initial members of the system are $\pi_{-1}=0$ and $\pi_0=1$. Then the formal solution of eq.~\eqref{system} can be expressed as
	\begin{equation}\label{conjecture}
	\varphi_n(t)=  i^n \pi_n(i\partial_t) C(t)\,,\qquad \varphi_0(t) = C(t)\,.
\end{equation}
Thanks to the Favard theorem  \cite{chihara}, the polynomials $\pi_n$ obeying the three-term recurrence relation (\ref{fav}) are orthogonal with respect to some weight (orthogonality measure) $\rho(x)$. We thus have
\begin{equation}\label{ort}
		\int_{-\infty}^\infty dx \rho(x) \pi_n(x) \pi_m(x) =  \delta_{n,m}\,,\qquad \int_{-\infty}^\infty dx \rho(x) =1\,.
\end{equation}
Requiring that the initial condition $\varphi_n(0)=\delta_{n,0}$ is satisfied for the solution (\ref{conjecture}) one can deduce a connection between $C(t)$ and the weight $\rho(x)$. Introducing the Fourier transform by $C(t) = \int_{-\infty}^\infty dx e^{-itx}\hat{C}(x)$, eq.~(\ref{conjecture}) becomes
	\begin{equation}\label{123}
		\varphi_n(t)= i^n  \int_{-\infty}^\infty dx e^{-itx}  \pi_n(x) \hat{C}(x).
	\end{equation}
Comparing the initial condition (\ref{123}) at $t=0$ with the orthogonality condition (\ref{ort}) taken at $m=0$, we infer that the weight corresponds to the Fourier transform of the autocorrelation function, $\rho(x)=\hat{C}(x)$. Therefore we obtain
	\begin{equation}\label{connection}
		C(t) =  \int_{-\infty}^\infty dx e^{-itx}\rho(x)\,.
	\end{equation}
Equation (\ref{connection}) completes the solution of eq.~(\ref{system}). Therefore, any set of orthogonal polynomials that satisfies the three-term recurrence relation (\ref{fav}) and obeys the orthogonality condition (\ref{ort}), provides one solution for the semi-infinite discrete Schr\"{o}dinger equation (\ref{system}). Since the sets of orthogonal polynomials are much widely studied than eq.~(\ref{system}), the latter knowledge is useful to explore eq.~(\ref{system}) and its consequences. As a simple example let us consider the classical Hermite polynomials $H_n(x)$. After proper rescaling, we obtain the polynomials $\pi_n(x)=H_n(x/\sqrt{2})/\sqrt{2^n n!}$ that satisfy eq.~\eqref{fav} with $b_n =\sqrt{n}$ (note that this does not conform with the universal operator growth hypothesis \eqref{UOGH}). Their weight is $\rho(x)=e^{-x^2/2}/\sqrt{2\pi}$. Using eq.~(\ref{123}) we then find the corresponding wave functions, $\varphi_n(t) = t^n e^{-t^2/2}/\sqrt{n!}$. The latter can also be checked by a direct substitution into eq.~\eqref{system}. 

Note that the wave functions $\varphi_n(t)$ together with the orthogonal polynomials $\pi_n(x)$ allow for a remarkable representation of the evolution operator. Indeed, let us consider an operator function
\begin{align}
F(t,\hat A)=\sum_{n=0}^{\infty}i^n\varphi_n(t)\pi_n(\hat A)\,.
\end{align}
Using the initial condition $\varphi_{-1}=\pi_{-1}=0$ it easy to show that $F(0,\hat A)=1$ and $\partial_t F(t,\hat A)=i\hat A F(t,\hat A)$. This is the first-order linear differential equation that has a unique solution
\begin{align}
F(t,\hat A)=e^{i t\hat A}\,.
\end{align}
Here $\hat A$ is an arbitrary time-independent operator. In the special case when $\hat A$ is the Liouville superoperator we thus obtain
\begin{equation}\label{soex}
    e^{i\;\!t\;\!\mathcal{L}} = \sum_{n=0}^\infty i^n\varphi_n(t) \pi_n\left(\mathcal{L}\right)\,.
\end{equation}
Equation (\ref{soex}) is the expansion of the evolution superoperator in terms of the polynomials, see for example refs.~\cite{TalEzer_1984_Accurate,Vijay1999,Chen_1999_Chebyshev,Weibe_2006_Kernel,Soares_2024_Non-unitary}. In fact, eq.~(\ref{soex}) can serve as a starting point for a systematic perturbative expansion, whose convergence can be made uniform in time by choosing a set of  $\varphi_n(t)$ with the long-time asymptotics fitting that of the actual autocorrelation function. This remark goes beyond the scope of the recursion method, highlighting the importance of orthogonal polynomials in a broader context of quantum dynamics.

\section{A family of models based on the continuous Hahn polynomials \label{sec: Hahn}}

\subsection{General model}

Equations~(\ref{bnsqrt}) and (\ref{mp}) provide an example \cite{Parker_2019} of the one-parameter family of the solutions of eq.~(\ref{system}) based on the Meixner--Pollaczek polynomials\footnote{While the Meixner--Pollaczek polynomials depend on two parameters, here we have in mind their one-parametric subclass with the weight given by the square of the gamma function.} \cite{Koekoek_2010}. Even though the polynomials are very involved, the wave functions (\ref{mp}) have a relatively simple form.

Here we generalize the latter one-parameter solution to a two-parameter solution of \eqref{system} with the linear growth of the corresponding Lanczos coefficients. Instead  of Meixner-Pollaczek polynomials, our solution is based on the continuous Hahn polynomials \cite{Koekoek_2010}. It has the Lanczos coefficients given by
\begin{equation}\label{bn1}
b^2_n = \frac{4n (n+2 a-1) (n+2 b-1) (n+2 a+2b-2)}{(2n+2 a+2 b-3) (2n+ 2 a+2 b-1)}\,,
\end{equation}
and the wave functions
\begin{equation}\label{ff1}
 \varphi_n(t) = \mathcal{F}_n \,\frac{\tanh^n(t)}{\cosh^{4a}(t)}\,\, \pFq{2}{1}{a-b+\frac{1}{2},2a+n}{a+b+n+\frac{1}{2}}{\tanh^2(t)}\,.
\end{equation}
Here the prefactor is given by\footnote{Note that a seeming similarity between $b_n$ and $b$ is accidental as they denote different objects.}
\begin{equation}\label{bn3}
    \mathcal{F}_n  = \frac{1}{n!} \left(\prod\limits_{k=1}^n b_k\right) = \sqrt{\frac{(2a)_n(2b)_n(2a+2b-1)_n}{n!(a+b-1/2)_n(a+b+1/2)_n}}\,,
\end{equation}
and
\begin{align}\label{eq:2F1}
\pFq{2}{1}{a,b}{c}{x}=\frac{\Gamma(c)} {\Gamma(a)\Gamma(b)}\sum_{k=0}^{\infty}\frac{\Gamma(a+k)\Gamma(b+k)}{\Gamma(c+k)}\frac{x^k}{k!}
\end{align}
is the Gauss hypergeometric function. The solution (\ref{ff1}) depends on two parameters $a$ and $b$ that can be either both positive real numbers
\begin{equation}\label{realreal}
a>0\,,\qquad  b>0\,,
\end{equation}
or a pair of complex conjugate numbers with a positive real part,
\begin{equation}\label{complexconjugate}
a=r+i\omega\,,\qquad  b=r-i\omega\,,\qquad r>0\,.
\end{equation}

The wave functions \eqref{ff1} have several equivalent representations that are discussed in appendix \ref{appendix}. Here we only list one such representation where the symmetry to the interchange of $a$ and $b$ is obvious,
\begin{equation}\label{wave2}
         \varphi_n(t) = \mathcal{F}_n\,\frac{\tanh^n(t)}{\cosh^{2a+2b-1}(t)} \, \pFq{2}{1}{a-b+\frac{1}{2},b-a+\frac{1}{2}}{a+b+n+\frac{1}{2}}{-\sinh ^2(t)}\,.
\end{equation}
The form (\ref{wave2}) manifestly demonstrates that $\varphi_n(t)$ is real for all admissible values of $a$ and  $b$ as all the summands in eq.~(\ref{eq:2F1}) are real for both choices (\ref{realreal}) and (\ref{complexconjugate}). Moreover, $\varphi_n(t)$ is positive at any $t$ for real and positive $a$ and $b$.  We note that the weight that corresponds to the exactly solvable sequence (\ref{bn1}) is given by
\begin{align}\label{measure}
\rho(x)=\frac{\Gamma(2a+2b)}{8\pi\Gamma(2a)\Gamma(2b)\Gamma(a+b)^2}| \Gamma(a+i x/4) \Gamma(b+i x/4)|^2\,.
\end{align}
Via eq.~(\ref{connection}) it determines $C(t)\equiv\varphi_0(t)$.

\subsection{Particular cases}

The exactly solvable sequence (\ref{bn1}) with the solution (\ref{ff1}) has two parameters and in special cases reduces to more elementary solutions that will be studied in the following. This simplification happens for the values of $a$ and $b$ for which the hypergeometric functions appearing in eqs.~(\ref{ff1}) or (\ref{wave2}) reduce to a simpler form, or for $a$ and $b$ for which the Fourier transform (\ref{connection}) of the weight (\ref{measure}) simplifies. For real parameters, the latter is achieved in the cases: (i) $a$ and $b$ are both positive integers, (ii) $a$ and $b$ are both odd half-integers, (iii) one parameter between $a$ and $b$ is a positive integer and the other is an odd half-integer, and (iv) the difference $b-a$ is an odd half-integer. There we can use the relations
\begin{gather}
|\Gamma(1+k+ix/4)|^2=\frac{\pi x/4}{\sinh(\pi x/4)}\prod_{j=1}^{k}(j^2+x^2/4)\,,\qquad k\in \mathbb{N}\,, \\
|\Gamma(1/2+k+ix/4)|^2=\frac{\pi}{\cosh(\pi x/4)}\prod_{j=1}^{k}((j-1/2)^2+x^2/4)\,,\qquad k\in \mathbb{N}\,,\\
\Gamma(z)\Gamma(z+1/2)=\sqrt\pi\, 2^{1-2z}\;\! \Gamma(2z)\,,
\end{gather}
in conjunction with $\Gamma(z+1)=z\;\!\Gamma(z)$ for a complex $z$ with a positive real part.

Let us work out several explicit results. Consider the case $b=a+k+1/2$ with a positive integer $k$. At $k=0$, for the choice $a=\eta/4$, using, e.g, the Ramanujan formula for the Fourier transform of the square of the Gamma function we obtain
\begin{align}\label{eq:MPMPMP}
	C(t)=\frac{1}{\cosh^\eta(t)}\,,\qquad b_n=\sqrt{n(n-1+\eta)}\,.
\end{align}
Moreover, in this case the hypergeometric function in eq.~(\ref{ff1}) becomes unity and we obtain the wave fuctions of eq.~(\ref{mp}). Equation (\ref{eq:MPMPMP}) corresponds to the solution based on the Meixner--Pollaczek polynomials, which was previously obtained by different means in ref.~\cite{Parker_2019}. 

For $k=1$ we have $b=a+3/2$ and

\begin{equation}
   C(t) = \frac{ a+\cosh ^2(t)}{(a+1)\cosh ^{4 a+2}(t)}\,,\qquad b_n^2 = \frac{n (2 a+n-1) (2 a+n+2) (4 a+n+1)}{(2 a+n) (2 a+n+1)}\,.
\end{equation}
Another elementary expression is obtained for the case $a=b=1/2$. It is given by
\begin{equation}\label{log1}
    C(t) = \frac{2t}{\sinh(2t)}\,,\qquad b_n^2 = \frac{4 n^4}{4 n^2-1}\,.
\end{equation}
We also notice that for $a=b=1/4$ the result can be expressed as
\begin{equation}\label{log2}
        C(t) = \frac{2 K\left(\tanh ^2(t)\right)}{\pi \cosh(t) }\,,\qquad b_n = n-\frac{1}{2}\,,
\end{equation}
where $K$ denotes the complete elliptic integral of the first kind. Notice an extremely simple form of $b_n$ in this case.

For the complex conjugated series, a notable example corresponds to $a=b^*=(1+i\omega)/2$. We find
\begin{equation}\label{bea123}
   C(t)= \frac{\sin(2\omega t)}{\omega \sinh(2t)}\,,\qquad b_n = \frac{4 n^2 \left(n^2+\omega ^2\right)}{4 n^2-1}\,.
\end{equation}
Therefore, the autocorrelation function shows the damped oscillations. At $\omega\to 0$, eq.~(\ref{bea123}) reduces to the result (\ref{log1}).

\subsection{Asymptotic behavior}

Let us study the asymptotic behavior of the autocorrelation function $C(t)$ at late times, $t\to\infty$. In the case of real parameters $b>a>0$, $C(t)$ can be obtained from eq.~(\ref{ff1}) by setting $\tanh(t)$ to $1$ in the hypergeometric function. Using the property\footnote{\url{https://functions.wolfram.com/07.23.03.0002.01}}
\begin{align}
\pFq{2}{1}{a,b}{c}{1}=\frac{\Gamma(c) \Gamma(c - a - b)}{\Gamma(c - a)\Gamma(c - b)}\,,\qquad \textrm{Re}(c - a - b) > 0\,,
\end{align}
we obtain
\begin{equation}\label{asymp 1}
C(t) \simeq \frac{\Gamma(2a+2b)\Gamma(b-a)}{\Gamma(a+b)\Gamma(2b)} e^{-4at}\,,\qquad b>a>0\,.
\end{equation}
The case $a>b>0$ follows from the symmetry as $C(t)$ is invariant to the exchange of $a$ and $b$. Finally the case $a=b$ is more difficult and can be obtained using the expansion\footnote{\url{https://functions.wolfram.com/07.23.06.0014.01}}
\begin{align}
\pFq{2}{1}{a,b}{a+b}{z}=\frac{\Gamma(a+b)}{\Gamma(a)\Gamma(b)} \left(\ln\frac{1}{1-z}+\frac{\Gamma'(a)}{\Gamma(a)}+\frac{\Gamma'(b)}{\Gamma(b)} +2\gamma_{\scriptscriptstyle{E}}\right)(1+O(1-z))\,,
\end{align}
where $z\to 1$ and $\gamma_{\scriptscriptstyle{E}}$ is the Euler constant. This yields
\begin{align}\label{logcase}
C(t)\simeq  \frac{4\;\!\Gamma(4a)}{[\Gamma(2a)]^2}\;\! t\;\! e^{-4 a t},
\end{align}
which is consistent with eqs.~(\ref{log1}) and (\ref{log2}). Equations (\ref{asymp 1}) and (\ref{logcase}) reveal that the autocorrelation function for the model (\ref{bn1}) decay as $C(t)\sim e^{-4\;\!\mathrm{min}(a,b)\;\!t+\delta_{a,b}\ln t}$. Note the logarithmic correction in the case $a=b$, which makes the decay slower.

In the complex-conjugate case $a=b^*$, we transform eq.~(\ref{ff1}) by making use of the expansion\footnote{\url{https://functions.wolfram.com/07.23.06.0008.01}}
\begin{align}
\pFq{2}{1}{a,b}{c}{z}=\left[\frac{\Gamma(c)\Gamma(a+b-c)}{\Gamma(a)\Gamma(b)} (1-z)^{c-a-b}+ \frac{\Gamma(c)\Gamma(c-a-b)}{\Gamma(c-a)\Gamma(c-b)} \right]&(1+O(1-z))\,,\notag\\
&c-a-b\notin \mathbb{Z}
\end{align}
where $z\to 1$. This gives
\begin{equation}\label{asymp 2}
C(t) \simeq   \frac{2^{4 r}}{\sqrt{\pi}}  \Gamma \left(2 r+\frac{1}{2}\right) {\mathrm Re}\left(\frac{\Gamma (2 i \omega ) e^{4 i  \omega t}}{\Gamma (2 r+2 i \omega )}\right)e^{-4 r t}\,,
\end{equation}
where the notation of eq.~\eqref{complexconjugate} is used. Note that at $\omega\to 0$, eq.~(\ref{asymp 2}) reduces to eq.~(\ref{logcase}), as it must be the case.

We eventually note that eq.~(\ref{asymp 2}) can also be expressed as
\begin{align}\label{asympfinal}
C(t) \simeq   \frac{\Gamma(2a+2b)}{\Gamma(a+b)}\left(\frac{\Gamma (a-b) e^{-4b t}}{\Gamma (2a)}+ \frac{\Gamma (b-a) e^{-4a t}}{\Gamma (2b)}\right)\,.
\end{align}
Equation (\ref{asympfinal}) is quite general and applies to both, complex and real cases. It reduces to the special cases (\ref{asymp 1}), (\ref{logcase}), and (\ref{asymp 2}) if the corresponding limits are taken.

\begin{figure}[htbp]
\centering
\includegraphics[width=.45\textwidth]{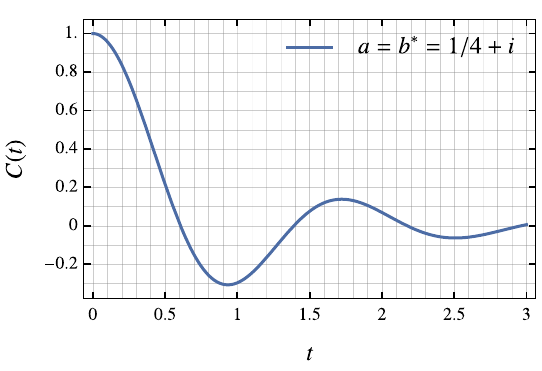}
\qquad
\includegraphics[width=.45\textwidth]{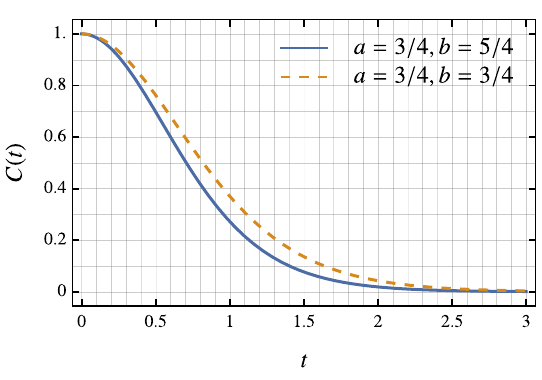}
\caption{Two typical behaviors of the autocorrelation function $C(t)=\varphi_0(t)$ given by eq.~\eqref{ff1}.   Depending on the choice of parameters indicated in the plots, the autocorrelation function can either feature damped oscillations (left plot) or damping without oscillations (right plot). We also see that the case $a=b=3/4$ shows slower relaxation from the one of $a=3/4$, $b=5/4$ due to an additional factor $t$ that multiplies the exponential decay in $C(t)$, see eq.~(\ref{logcase}).
\label{Fig1}}
\end{figure}

From the asymptotic expansions (\ref{asymp 1}) and (\ref{logcase}) we note that the decay of $C(t)$ is monotonic for real parameters $a$ and $b$. This should be contrasted to the case of complex parameters (\ref{asymp 2}), where the damped oscillations occur.
The typical behavior of the autocorrelation function is shown in figure~\ref{Fig1}.

Let us connect the late-time asymptotic behavior of $C(t)$ with the asymptotic properties of the Lanczos coefficients $b_n$. Expansion of $b_n$ of eq.~(\ref{bn1}) at $n\to \infty$ is given by
\begin{equation}
b_n = n + a + b -1 - \frac{(a+b-1)^2+(a-b)^2-1/4}{2n}+ O(1/n^2)\,.
\end{equation}
We can see that if an asymptotic expansion of a generic $b_n$ with linear growth is denoted as $b_n=n+b^{(1)}+ b^{(2)}n^{-1} + o(n^{-1})$, then the solution \eqref{ff1} can be used to model a system with the subleading coefficients $b^{(1)}>-1$ and arbitrary $b^{(2)}$. To discriminate the two cases of $\omega=0$ and $\omega\neq 0$ in terms of $b^{(1)}$ and $b^{(2)}$, one needs to compute the sign of
\begin{equation}\label{crit}
    \frac{1}{8}-\frac{( b^{(1)} )^2}{2}-b^{(2)}\,.
\end{equation}
It is positive whenever  the parameters $a$ and $b$ are real and $a\neq b$, corresponding to $\omega=0$. Conversely, the expression (\ref{crit}) is negative if $a$ and $b$ are complex-conjugate pairs, which corresponds to $\omega\neq 0$. Finally the expression (\ref{crit}) nullifies if $a=b$. In this case there is an additional $t$ factor in $C(t)$, see eq.~(\ref{logcase}).

\subsection{Derivation based on the properties of the hypergeometric functions}

Detailed construction of the functions \eqref{ff1} [or equivalently (\ref{wave2})] that solve eq.~\eqref{system} with the Lanczos coefficients \eqref{bn1} and their relation to the continuous Hahn polynomials is given in appendix \ref{appendix}. Here we show that eq.~\eqref{wave2} solves the system \eqref{system} directly by making use of some basic properties of the hypergeometric functions. Substituting eq.~\eqref{wave2} into eq.~\eqref{system}, after the change of variables we obtain
\begin{equation}
    \left(n + x (2a+2b-1)+2x(1-x)\frac{d}{dx}\right)\Phi_n(x) = n(1-x)\Phi_{n-1}(x)+ \frac{b_{n+1}^2}{n+1} x\;\! \Phi_{n+1}(x)\,,
\end{equation}
where we have defined
\begin{equation}
\Phi_n(x)  = \pFq{2}{1}{a-b+\frac{1}{2}, b-a+\frac{1}{2}}{ a+b+n+\frac{1}{2}}{x}\,.
\end{equation}
The derivative can be evaluated by means of the property\footnote{\url{http://dlmf.nist.gov/15.5.E21}}
\begin{align}
\left[c-a-b+(1-z)\frac{d}{dz}\right]\, \pFq{2}{1}{a,b}{c}{z}=\frac{(c-a)(c-b)}{c} \pFq{2}{1}{a,b}{c+1}{z}\,,
\end{align}
leading to
\begin{equation}
\left(1-2x\right)\Phi_n(x) = (1-x)\Phi_{n-1}(x) -\frac{4(2a+n)(2b+n)}{4(a+b+n)^2-1} x\;\!\Phi_{n+1}(x)\,.
\end{equation}
The latter relation holds as one of the contiguous properties for the hypergeometric functions\footnote{\url{http://dlmf.nist.gov/15.5.E18}}. We have therefore shown that the functions \eqref{wave2} [or equivalently (\ref{ff1})] are the solution of eq.~\eqref{system}.

\subsection{Krylov complexity}

The spread of the operator dynamics can be quantified by the Krylov complexity  ${\cal K} (t)$. It is defined by \cite{Parker_2019}
\begin{equation}
    \mathcal{K}(t) = \sum_{n=0}^\infty n \varphi_n(t)^2 = \frac{d \mathcal{K}{_\lambda}(t) }{d\lambda}\bigg|_{\lambda=1}\,,\qquad \mathcal{K}_{\lambda}(t) \equiv \sum_{n=0}^\infty \lambda^n \varphi_n(t)^2 \,.
\end{equation}
Using an integral representation for the hypergeometric function\footnote{\url{https://dlmf.nist.gov/15.6.E1}} given by
\begin{align}
\pFq{2}{1}{a-b+\frac{1}{2},2 a+n}{a+b+n+\frac{1}{2}}{T}= \frac{\Gamma\left(a+b+n+\frac{1}{2}\right)}{\Gamma(2a+n) \Gamma\left(b-a+\frac{1}{2}\right)} \int_0^1 du \frac{u^{2a+n-1}}{ [(1-u)(1-Tu)]^{a-b+1/2}}\,,
\end{align}
where $\mathrm{Re}(b-a)>-1/2$ is assumed, we can represent
\begin{align}\label{been1}
\mathcal{K}_\lambda(t)={}& \frac{(1-T)^{2a+2b}\cos (\pi  (a-b)) \Gamma \left(a+b+\frac{1}{2}\right)^2}{\pi  \Gamma (2 a) \Gamma (2 b)} \notag\\
&\times\int_0^1\int_0^1 \frac{1+T\lambda uv}{(1-T\lambda uv)^{2a+2b}}  \frac{\left(\frac{(1-u)(1-uT)}{(1-v)(1-vT)}\right)^{b-a}u^{2a-1}v^{2b-1}du dv}{\sqrt{(1-u)(1-uT)(1-v)(1-vT)}}\,.
\end{align}
Here $T=\tanh^2(t)$ and $-1/2<\mathrm{Re}(b-a)<1/2$. The normalization condition $\mathcal{K}_1(t)=1$ enables us to infer the integral
\begin{align}\label{n1}
&\int_0^1\int_0^1 \frac{1+T uv}{(1-T uv)^{2a+2b}} \frac{\left(\frac{(1-u)(1-uT)}{(1-v)(1-vT)}\right)^{b-a}  u^{2a-1}v^{2b-1}du dv}{\sqrt{(1-u)(1-uT)(1-v)(1-vT)}}\notag\\
&= \frac{\pi  \Gamma (2 a) \Gamma (2 b)}{(1-T)^{2a+2b}\cos (\pi  (a-b)) \Gamma \left(a+b+\frac{1}{2}\right)^2}\,.
\end{align}
The complexity can be obtained by differentiating eq.~(\ref{been1}) and reads
\begin{align}\label{mrbean}
 \mathcal{K}(t) ={}&  \frac{T(1-T)^{2a+2b}\cos (\pi  (a-b)) \Gamma \left(a+c+\frac{1}{2}\right)^2}{\pi  \Gamma (2 a) \Gamma (2 c)}\notag\\
&\times\int_0^1\int_0^1 \frac{2 (a + b) (1 + T u v) + (1 - T u v)}{(1-T uv)^{2a+2b+1}}  \frac{\left(\frac{(1-u)(1-uT)}{(1-v)(1-vT)}\right)^{b-a} u^{2a}v^{2b}du dv}{\sqrt{(1-u)(1-uT)(1-v)(1-vT)}}\,.
\end{align}
This formula reduces  to  $ \mathcal{K}(t)=\eta\,\sinh^2 (t)$ in the special case \eqref{bnsqrt} \cite{Bhattacharjee_2023_Operator}.\footnote{Notice that this case corresponds to $b=a+1/2$, so one has to understand the complexity using the limit \[
\lim\limits_{b\to a+1/2} \cos (\pi (b-a))\int\limits_0^1 \frac{f(v) }{(1-v)^{b-a+1/2}} dv = \pi f(1)\,.\]} 

Note that eq.~(\ref{been1}) allows one to also address the generalized complexity \cite{fan2023generalised} $K^{(\delta)} = (\lambda\partial_\lambda)^\delta K_\lambda \Big|_{\lambda=1}$ for positive integers $\delta$. At the same time, another widespread measure of complexity, the Krylov entropy \cite{Barbon_2019_On_evolution}, can not be obtained in an analogous way.

Let us compute the Krylov complexity (\ref{mrbean}) at late times $t\to+\infty$ for which it suffices to consider the leading order as $T\to 1$. In this case, it is obvious that the main contribution to the integrals comes from the regions $u\sim 1$ and $v\sim1$ and therefore in the leading order we can approximate
    \begin{align}\nonumber
   \mathcal{K}(t) \simeq {}& \frac{T(1-T)^{2a+2b}\cos (\pi  (a-b)) \Gamma \left(a+b+\frac{1}{2}\right)^2}{\pi  \Gamma (2 a) \Gamma (2 b)}2 (a + b)\\
    &\times\int_0^1\int_0^1 \frac{1+T u v }{(1-T uv)^{2a+2b+1}}  \frac{\left(\frac{(1-u)(1-uT)}{(1-v)(1-vT)}\right)^{b-a} du dv}{\sqrt{(1-u)(1-uT)(1-v)(1-vT)}}\,.
\end{align}
Within the same accuracy, we can insert the appropriate powers of $u$ and $v$, to get the integral identical to eq. \eqref{n1} up to shifting  $a$ and $b$ by $1/4$.
In this way as $t \to +\infty$ we obtain
\begin{equation}\label{K asymp}
    \mathcal{K}(t) \simeq k_\infty e^{2t}\,,\qquad k_\infty = \frac{ \Gamma \left(2 a+\frac{1}{2}\right) \Gamma \left(2 b+\frac{1}{2}\right) \Gamma \left(a+b+\frac{1}{2}\right)^2}{2\;\!\Gamma (2 a) \Gamma (2 b) \Gamma (a+b) \Gamma (a+b+1)}\,.
\end{equation}
The exponential growth of the complexity is consistent with the universal operator growth~\cite{Parker_2019}. Although the result (\ref{K asymp}) is derived under restricted values for $a$ and $b$, it turns out that the latter restriction is irrelevant for eq.~(\ref{K asymp}) that remains valid for all admissible $a$ and $b$. We have confirmed this by a numerical analysis.

\section{A family of models with odd-even alterations in the Lanczos sequence \label{sec: alternating}}

In the previous section we have encountered examples of the oscillating behavior in the autocorrelation function. Now we will show that there exists another way to model such behavior of the autocorrelation function. It is based on an odd-even alteration in the subleading terms of $b_n$. Let us consider a very simple $C(t)$ of the form
\begin{equation}\label{Ali}
    C(t) = \frac{\cos(\omega t)}{\cosh(t)}\,.
\end{equation}
We will demonstrate that the corresponding Lanczos coefficients that lead to eq.~(\ref{Ali}) are given by
\begin{equation}\label{bn2}
b_n =
\begin{cases}
n,  & n \in 2\mathbb{Z}\\
\sqrt{n^2+\omega^2}, & n \in 2\mathbb{Z}+1
\end{cases}.
\end{equation}
Note that now we are now solving the inverse problem for eq.~(\ref{system}). For a given $\varphi_0(t)\equiv C(t)$ we seek the coefficients $b_n$ such that eq.~(\ref{system}) is satisfied subject to the initial condition (\ref{eq:bc}). In principle, the solution can be obtained as follows. Once $C(t)$ is known, the weight for $\pi$-polynomials can be found via the Fourier transform (\ref{connection}). Knowing the weight, one can construct the orthogonalized set of $\pi$-polynomials that should satisfy the three-term recurrence relation (\ref{fav}) that in turn determines $b_n$'s. We will, however, approach this problem differently elaborating upon the method of moments \cite{chihara}.

It is convenient to introduce monic polynomials $P_n\equiv \pi_n \prod_{k=1}^n b_k$, so instead of the relation \eqref{fav} we have
\begin{equation}\label{rec1}
	x P_{n}(x) =  P_{n+1}(x) + b_n^2 P_{n-1}(x).
\end{equation}
These polynomials are orthogonal with respect to the measure $\rho(x)$,
\begin{equation}\label{ortP}
\int_{-\infty}^\infty dx \rho(x) P_n(x) P_m(x) = h_n \delta_{n,m}\,,\qquad h_n>0\,.
\end{equation}
Let us recall how $\rho(x)$ (or $C(t)$) and $b_n^2$ are connected.
Integrating both sides of eq.~\eqref{rec1} multiplied by $\rho(x) P_{n-1}(x)$ leads to the connection
$
    h_n = b_n^2 h_{n-1}.
$
On the other hand, we can define the partition function $Z_n$, which takes the form of a Hankel determinant
\begin{equation}\label{hankel}
    Z_n = \frac{1}{n!} \int_{-\infty}^\infty dx_1\rho(x_1)  \int_{-\infty}^\infty dx_2\rho(x_2) \cdots \int_{-\infty}^\infty dx_n\rho(x_n) \det\limits_{1\le i,j\le n}(x_i^{j-1})^2 = \det\limits_{0\le i,j \le n-1} \mu_{i+j}\,.
\end{equation}
Here the moments are given by
\begin{equation}\label{moments}
\mu_k = \int_{-\infty}^\infty dx \rho(x) x^k  = (i\partial_{t})^k C(t) \Big|_{t=0}\,,
\end{equation}
see eq.~(\ref{connection}) for the second equality. By making use of the identity
\begin{align}
\det\limits_{1\le i,j\le n}(x_i^{j-1})=\det\limits_{1\le i,j\le n}(P_{j-1}(x_i))
\end{align}
that holds for monic polynomials (we can manipulate with the columns of the matrix on the left-hand side and form polynomials without changing the value of the determinant), we can express $Z_n$ also as
\begin{equation}
    Z_n = \frac{1}{n!} \int_{-\infty}^\infty dx_1\rho(x_1)  \int_{-\infty}^\infty dx_2\rho(x_2) \cdots \int_{-\infty}^\infty dx_n\rho(x_n)  \det_{1\le i,j\le n}(P_{j-1}(x_i))^2  = \prod\limits_{j=0}^{n-1}h_j\, ,
\end{equation}
where we have used the orthogonality (\ref{ortP}). Therefore we obtain
\begin{equation}\label{bZ}
    b_n^2 = \frac{Z_{n+1}Z_{n-1}}{Z_n^2}\,.
\end{equation}
Equation (\ref{bZ}) should be understood as the connection between $b_n$'s and the moments $\mu_k$ encoded into the Hankel determinant (\ref{hankel}).

Further simplification to the problem can be achieved if the measure is symmetric, $\rho(x) = \rho(-x)$.
In this case following ref.~\cite{Clarkson2023} we can express
\begin{equation}
    Z_{2n} = A_n B_n\,,\qquad Z_{2n+1} =  A_{n+1} B_n\,,
\end{equation}
where the \textit{partial} partition functions are defined by
\begin{equation}\label{ab}
    A_n = \det_{0\le i,j \le n-1} \mu_{2i+2j}\,,\qquad B_n = \det_{0\le i,j \le n-1} \mu_{2i+2j+2}\,.
\end{equation}
It then follows
\begin{equation}\label{zzz}
b_{2n}^2 = \frac{A_{n+1}B_{n-1}}{A_{n}B_n}\,,\qquad
b_{2n+1}^2 = \frac{A_nB_{n+1}}{A_{n+1}B_n}\,.
\end{equation}

Interestingly, the moments \eqref{moments} corresponding to $C(t)$ of eq.~(\ref{Ali}) can be expressed in terms of the Euler polynomials $E_n(x)$, which are defined as
\begin{equation}
    \frac{2e^{xt}}{e^t+1} =\sum\limits_{n=0}^\infty E_{n}(x) \frac{t^n}{n!}.
\end{equation}
Indeed, we find $\mu_{2k+1}=0$ and
\begin{equation}
    \mu_{2k} =  (-4)^k  E_{2k}\left(\frac{1+i\omega}{2}\right).
\end{equation}
Therefore,
\begin{equation}\label{mu2}
\det_{0\le i,j \le n-1} \mu_{2i+2j+2\eta} =4^{n(n-1)}(-4)^{n\eta} \det_{0\le i,j \le n-1} E_{2i+2j+2 \eta}\left(\frac{1+i\omega}{2}\right).
\end{equation}
The expressions for the type of Hankel determinants appearing in eq.~(\ref{mu2}) for $\eta = 0$ and $\eta = 1$ are available in the literature \cite{dets}. In particular, using corollary 5.2 of ref.~\cite{dets} we obtain
\begin{equation}\label{hdet}
\det_{0\le i,j \le n-1} \mu_{2i+2j+2\eta} = \left((-4)^\eta E_{2\eta}\left(\frac{1+i\omega}{2}\right)\right)^n\times \prod\limits_{\ell=1}^{n-1} c_{\eta,\ell}^{n-\ell}\,,
\end{equation}
with $c_{\eta,\ell} = 4\ell^2 [\omega^2 + (2\ell+2\eta-1)^2]$. Equation (\ref{hdet}) enables us to conclude
\begin{equation}
    \frac{A_{k+1}}{A_k} = \prod\limits_{\ell =1}^k c_{0,\ell}\,,\qquad  \frac{B_{k+1}}{B_k} = (1+\omega^2)\prod\limits_{\ell =1}^k c_{1,\ell}\,.
\end{equation}
We therefore obtain
\begin{equation}
b_{2k}^2 = \frac{A_{k+1}B_{k-1}}{A_{k}B_k} =   \frac{c_{0,1}}{1+\omega^2} \prod\limits_{\ell =1}^{k-1} \frac{c_{0,\ell+1}}{c_{1,\ell}} = 4 \prod\limits_{\ell =1}^{k-1}  \frac{(\ell+1)^2}{\ell^2} = (2k)^2,
\end{equation}
\begin{equation}
    b_{2k+1}^2 =  \frac{A_kB_{k+1}}{A_{k+1}B_k}=(1+\omega^2) \prod\limits_{\ell =1}^{k} \frac{c_{1,\ell}}{c_{0,\ell}}  = (1+\omega^2) \prod\limits_{\ell =1}^{k}
    \frac{\omega^2+(2\ell+1)^2}{\omega^2+(2\ell-1)^2} = \omega^2 + (2k+1)^2\,.
\end{equation}
This finishes the proof of eq.~\eqref{bn2}.

\section{Correlation functions relaxing to nonzero stationary values \label{sec: nonzero equilibrium}}

\subsection{Lanczos coefficients from stationary value}

In all of the previously considered examples, the autocorrelation function decays to zero at late times. However, in general this is not the case -- the late-time stationary value of the autocorrelation function can be nonzero. Here we explore the implications this incurs to the Lanczos coefficients.

Let us consider an autocorrelation function $C(t)$ that decays to zero. We study its deformation according to
 \begin{equation}\label{deform}
C^{(\kappa)}(t) = \kappa + (1-\kappa)C(t)\,,\qquad 0<\kappa<1\,,
\end{equation}
such that
\begin{equation}\label{limittt}
C^{(\kappa)}(t)\xrightarrow{t\rightarrow\infty} \kappa\,.
\end{equation}
We aim to find the changes in the Lanczos coefficients produced by the $\kappa$-deformation (\ref{deform}) of the autocorrelation function.

 Note that positive $\kappa$ is consistent with physical behavior of the autocorrelation function. Indeed,
\begin{align}
C^{(\kappa)}(t)\equiv{\rm Tr} \{ \mathcal{O}(t)\,\mathcal{O}\}/{\rm Tr}\{ \mathcal{O}^2\} =\sum_{E,E'} \langle E|\mathcal{O}|E'\rangle\langle E'| \mathcal{O}|E\rangle e^{-it(E'-E)} /{\rm Tr}\{ \mathcal{O}^2\}\,,
\end{align}
where $E,E'$ and $|E\rangle,\,|E'\rangle$ are the eigenenergies and the eigenvectors, respectively. If there are degeneracies in the spectrum, we choose the eigenvectors in such a way that $\mathcal{O}$ is diagonal in invariant subspaces.  All oscillating exponents are canceled with the help of the identity $\kappa=\lim_{t\rightarrow\infty} t^{-1}\int_0^t dt' C^{(\kappa)}(t') $. Note that the latter limit exists even in those pathological cases when the limit (\ref{limittt}) does not exist, in which cases it can be regarded as a definition of $\kappa$. Combining the above equalities, we get  $\kappa=\sum_{E} | \langle E|\mathcal{O}|E\rangle|^2 /{\rm Tr}\{ \mathcal{O}^2\}\geq0$.

Using eq.~\eqref{moments} one can immediately conclude that the moments corresponding to $C^{(\kappa)}(t)$ are given by
\begin{equation}\label{newmom}
   \mu^{(\kappa)} _0 = \mu_0=1\,,\qquad  \mu_{2k}^{(\kappa)} = (1-\kappa)\mu_{2k}\,,\qquad k>0\,,
\end{equation}
where $\mu_{2k}$ are the moments corresponding to $C(t)$. The corresponding deformations for the partial partition functions \eqref{ab} are given by
\begin{equation}\label{shifted}
    B_n^{(\kappa)} = (1-\kappa)^n B_n\,,\qquad A_n^{(\kappa)}  =  (1-\kappa)^n A_n + \kappa (1-\kappa)^{n-1} C_{n-1}\,,
\end{equation}
where
\begin{equation}\label{C}
C_n = \det_{0\le i,j \le n-1} \mu_{2i+2j+4}\,.
\end{equation}
Note that the determinant (\ref{C}) has not been encountered before. As there is no obvious way to relate $C_n$ to $A_n$ and $B_n$ of eq.~(\ref{ab}), it is thus not clear how to proceed further from this point following the recipe of the previous section.

In order to circumvent the latter obstacle, we use the moment-generating function instead, $G(z) = \sum_{k=0}^\infty \frac{\mu_{k}}{z^{k+1}}$.
Since we assumed that $C(t)$ is symmetric and thus zero odd moments, $
\mu_{2k+1}=0$, we have
\begin{equation}
    G(z) =\frac{1}{z} + \sum\limits_{k=1}^\infty \frac{\mu_{2k}}{z^{2k+1}}\,.
\end{equation}
A property of the generating function is that it encodes all the Lanczos coefficients $b_n$ once it is presented as a continued fraction  \cite{chihara,Parker_2019},
\begin{equation}\label{G}
    G(z)={\cfrac {1}{z-{\cfrac {b_1^2}{z-{\cfrac {b_2^2}{z-{\cfrac {b_3^2}{z-\dots}}}}}}}}\,.
\end{equation}

We now consider the $\kappa$-deformed case. Taking into account the changes of the moments described by eq.~(\ref{newmom}), we easily find that the deformation of the generating function is given by
\begin{equation}
    G^{(\kappa)}(z) = \frac{\kappa}{z} + (1 -\kappa) G(z)\,.
\end{equation}
Accounting for the continued fraction representation of $G^{(\kappa)}(z)$ similar to eq.~(\ref{G}), we can calculate the corresponding Lanczos coefficients. The final result reads
\begin{equation}\label{bkappa}
\left( \frac{b^{(\kappa)}_{2n}}{b_{2n}}\right)^2 =
      \frac{1 + \frac{\kappa}{1-\kappa} \delta_{n+1}} {1 + \frac{\kappa}{1-\kappa} \delta_n}\,,\qquad    \left( \frac{b^{(\kappa)}_{2n+1}}{b_{2n+1}}\right)^2 =
      \frac{1 + \frac{\kappa}{1-\kappa} \delta_{n}} {1 + \frac{\kappa}{1-\kappa} \delta_{n+1}}\,,
\end{equation}
where $\delta_0=0$, $\delta_1 =1$, $\delta_2= 1+ b^2_1/b^2_2$, $\delta_3= \delta_2 + b^2_1b^2_3/b^2_2b^2_4$, etc. The general expression is given by
\begin{equation}\label{delta}
\delta_{n} = 1+ \sum_{k=1}^{n-1} y_k\,, \qquad y_k= \prod_{j=1}^k \left(\frac{b_{2j-1}}{b_{2j}}\right)^2\,,
\end{equation}
where $n=2,3,4\ldots$. Equations (\ref{bkappa}) and (\ref{delta}) determine the Lanczos coefficients for the $\kappa$-deformed case characterized by the autocorrelation function (\ref{deform}).

It is instructive to derive an asymptotic form of eq.~\eqref{bkappa} in the case the Lanczos coefficients satisfy the universal operator growth hypothesis (\ref{UOGH}). As long as $b_n = n + O(1)$,  we obtain $y_k=c_0/k$ at the leading order, where the coefficient $c_0$ depends on the first subleading term in $b_n$ and thus $\delta_n = c_0\ln n + O(1)$. In this way we obtain
\begin{gather}
b^{(\kappa)}_{2n} = b_{2n}  + \frac{1}{\ln (g(\kappa)n)}+\cdots\,,\\
b^{(\kappa)}_{2n+1} = b_{2n+1} - \frac{1}{\ln (g(\kappa)n)}+\cdots\,,
\end{gather}
where $g(\kappa)$ depends on the subleading terms and the ellipsis denotes further subleading terms. One can see that the nonzero stationary value of the autocorrelation function is ensured by the odd-even alterations in the Lanszos coefficients that decay as the inverse logarithm of $n$, which is slower than any power law. While it has been known that a nonzero asymptotic value of the correlation function is associated with an alternating term in the Lanczos coefficients \cite{Viswanath_1994_Ordering, Yates_2020_Lifetime,Yates_2020_Dynamics, Bhattacharjee_2022_Krylov,Avdoshkin_2024_Krylov,Uskov_Lychkovskiy_2024_Quantum,dodelson2025black}, the precise scaling of the alternating term has not been derived. Note that the inverse logarithmic scaling has been considered among other possibilities in refs.~ \cite{dodelson2025black,Bhattacharjee_2022_Krylov}.

It is interesting that the \textit{shifted} determinant $C_n$ in eq.~\eqref{C} can be expressed via $A_n$ of eq.~(\ref{ab}).
Indeed, taking into account eqs.~\eqref{shifted} and (\ref{zzz}) we obtain
\begin{equation}
    \left( \frac{b^{(\kappa)}_{2n}}{b_{2n}}\right)^2 =
      \frac{1 + \frac{\kappa}{1-\kappa} \frac{C_{n}}{A_{n+1}}} {1 +\frac{\kappa}{1-\kappa} \frac{C_{n-1}}{A_{n}}}\,.
\end{equation}
Comparing this with eq.~\eqref{bkappa} we conclude that
\begin{equation}\label{3}
    C_n = A_{n+1} \delta_{n+1}\,.
\end{equation}

Let us study an example. For $b_n$ of eq.~\eqref{bn2} from the previous section, we obtain
\begin{equation}
    y_k = \frac{\Gamma \left(k+\frac{1}{2} -\frac{i \omega }{2}\right) \Gamma \left(k+\frac{1}{2}+\frac{i \omega }{2}\right)}{(k\;\!!)^2\, \Gamma \left(\frac{1}{2}-\frac{i \omega }{2}\right) \Gamma \left(\frac{1}{2}+ \frac{i \omega }{2}\right)}\,.
\end{equation}
Accounting for the determinant \eqref{mu2}, eq.~\eqref{3} then leads to
\begin{equation}\label{newhankel}
    \frac{ \det_{0\le i,j \le n-1} E_{2i+2j+4}(z)}{ \det_{0\le i,j \le n} E_{2i+2j}(z)} = 1+\frac{\sin(\pi z)}{\pi}\sum_{k=1}^n  \frac{\Gamma (k+1-z) \Gamma (k+z)}{(k\;\!!)^2}\,,
\end{equation}
where we have introduced $z=(1+i\omega)/2$. Note, however, that eq.~(\ref{newhankel}) is valid at any complex number $z$. We should also note that the two Hankel determinants of Euler polynomials in the left-hand side of eq.~(\ref{newhankel}) have different dimensions. The one known in the literature \cite{dets} is given by
\begin{align}
\det_{0\le i,j \le n} E_{2i+2j}(z)=(-1)^{\frac{n(n+1)}{2}} \prod_{\ell=0}^{n}\left((\ell+1)^2 (z+\ell)(z-\ell-1)\right)^{n-\ell}\,,
\end{align}
and the other one, $\det_{0\le i,j \le n-1} E_{2i+2j+4}(z)$, is calculated here in  eq.~(\ref{newhankel}).

The problem studied in this section has yet another interpretation. We assumed that the coefficients $b_n$ correspond to the autocorrelation function $C(t)$, which is related to the polynomials that satisfy the three-term recurrence relation (\ref{rec1}) and the orthogonalization condition (\ref{ortP}) with $\rho(x)$ as the weight. The obtained coefficients $b_n^{(\kappa)}$ also correspond to a set of orthogonal polynomials with the three-term recurrence relation and the orthogonalization condition, but with respect to the weight
\begin{align}\label{newweight}
\rho^{(\kappa)}(x)=(1-\kappa)\rho(x)+\kappa \delta(x).
\end{align}
We have therefore found the set of coefficients $b_n^{(\kappa)}$  for the weight (\ref{newweight}) knowing the set of coefficients $b_n$ for the weight $\rho(x)$.

\subsection{The stationary value from the Lanczos coefficients}

Let us consider the reversed problem that consists of finding the stationary, late-time, value $\kappa$ of the autocorrelation function $C^{(\kappa)}(t)$ provided the Lanczos coefficient $b^{(\kappa)}_n$ are known. To achieve that let us consider
\begin{align}
y_k^{(\kappa)}= \prod_{j=1}^k \left(\frac{b_{2j-1}^{(\kappa)}}{b_{2j}^{(\kappa)}}\right)^2\,.
\end{align}
Using eq.~(\ref{bkappa}) we find the connection
\begin{align}
y_k=(1-\kappa)\left(1+\frac{\kappa}{1-\kappa}\delta_k\right) \left(1+\frac{\kappa}{1-\kappa}\delta_{k+1}\right) y_k^{(\kappa)}\,.
\end{align}
On the other hand, the left-hand side can be expressed as $y_k=\delta_{k+1}-\delta_k$, as follows from  eq.~(\ref{delta}). We thus obtain
\begin{align}
y_k^{(\kappa)}=\frac{1}{\kappa}\left( \frac{1}{1+\frac{\kappa}{1-\kappa}\delta_k}- \frac{1}{1+\frac{\kappa}{1-\kappa}\delta_{k+1}}\right)\,.
\end{align}
For $\delta_k$ that diverges logarithmically with $k$ (or any other $\delta_k$ that tends to infinity as $k\to\infty$), we therefore obtain
\begin{equation}\label{kappa}
\kappa=1/\delta^{(\kappa)}_\infty =\left(1+\sum\limits_{k=1}^\infty y_k^{(\kappa)}\right)^{-1}\,.
\end{equation}
In the case of a divergent sum in the right-hand side of eq.~(\ref{kappa}), we obtain $\kappa=0$. This is consistent with the scaling $y_k \sim 1/k$ obtained previously for the undeformed $C(t)$ that decays to zero.

We find it instructive to also outline a {\it wrong} derivation of the stationary value from the Lanczos coefficients. Equation~\eqref{system} involves the sum rule
\begin{equation} \label{sum rule}
\sum\limits_{n=0}^\infty\varphi_n(t)^2 =1\,.
\end{equation}
If the convergence of $\varphi_n(t)$ to the corresponding stationary values $\varphi_n(\infty)$ (where $\varphi_0(\infty)\equiv \kappa$) were uniform, then the exchange of the limit $t\to\infty$ and the sum would give the sum rule   $\sum_{n=0}^\infty\varphi_n(\infty)^2 =1$. At the same time, eq.~\eqref{system} implies  $\varphi_{2l+1}(\infty)=0$ and $\varphi_{2n}(\infty)=\big(b_{2n-1}^{(\kappa)}/b_{2n}^{(\kappa)}\big)^2\, \varphi_{2n-2}(\infty)$, which leads to  $\varphi_{2n}^2=\varphi^2_0 y_n^{(\kappa)}$\;\!. Substituting this into the sum rule leads to $\kappa=\left(1 +  \sum_{k=1}^\infty y_k^{(\kappa)}\right)^{-1/2}$, which differs from the correct expression \eqref{kappa}.

This apparent paradox is resolved by noting that the convergence of $\varphi_n(t)$ is not, in general, uniform: at any fixed $t$ an infinite number of $\varphi_n(t)$ with sufficiently large $n$ are not anything close to their stationary values $\varphi_n(\infty)$. Therefore, in general, we cannot exchange the limit $t\to\infty$ and the infinite summation (that contains another limit). One can, however, derive a relaxed version of the sum rule by taking the limit of infinite time in the inequality that involves a sum of finite number of terms, $\sum_{n=0}^p\varphi_n(t)^2 \leq 1 $, where $p$ is finite. This way one obtains
\begin{equation}
\sum\limits_{n=0}^\infty\varphi_n(\infty)^2\leq 1.
\end{equation}
One can easily verify that this inequality does not lead to any contradiction with eq.~\eqref{kappa}.

The sum rule \eqref{sum rule} is widely used \cite{Barbon_2019_On_evolution,Muck_2022_Krylov,Bhattacharjee_2022_Krylov}. The above apparent paradox and its resolution highlights that it should be treated with caution at asymptotically large times.

\section{Discussion \label{sec: discussion}}

Our findings highlight the pivotal role of the subleading terms of the Lanczos coefficients $b_n$ at large $n$. In the exactly solvable two-parameter family \eqref{bn1} one can separately control two subleading terms proportional to $O(1)$ and $O(1/n)$ by tuning the parameters $a$ and $b$. Asymptotic analysis of the exact solution (see eqs.~\eqref{asymp 1} and \eqref{asymp 2}) shows that even the most rough feature of the correlation function -- its decay exponent -- depends on {\it both} subleading terms. Furthermore, subleading terms determine whether the correlation function features damped oscillations or damping without oscillations at large times, as illustrated in Fig. \ref{Fig1}. In contrast, the Krylov complexity  turns out to be largely  insensitive to the subleading terms, see eq.~\eqref{K asymp}.

The second family we studied, see eq.~\eqref{bn2}, features an alternating subleading term $O\big((-1)^n/n\big)$. This term provides a different pathway to damped oscillations in the autocorrelation function.

Finally, we have shown how Lanczos coefficients should be modified to obtain a deformation of the correlation function with a nonzero stationary value.

Importantly, the strong effect of subleading terms calls for caution when using  techniques that rely on approximation and/or extrapolation  of Lanczos coefficients. This includes various extrapolated versions of the recursion method \cite{viswanath2008recursion,Parker_2019, Uskov_Lychkovskiy_2024_Quantum,Wang_2024_Diffusion,Teretenkov_2025}, approximating Lanczos coefficients by sampling \cite{De_2024_Stochastic} and continuum approximation \cite{Parker_2019,Barbon_2019_On_evolution,Yates_2020_Dynamics,Muck_2022_Krylov,Bhattacharjee_2022_Krylov}.

The latter technique deserves a separate remark. In the continuum approximation, one treats the Lanczos coefficients $b_n=b(n)$ as a continuous function of $n$ \cite{Parker_2019,Barbon_2019_On_evolution,Muck_2022_Krylov}. In the presence of odd-even alterations, one treats $b_{2n-1}$ and $b_{2n}$ as two different continuous functions \cite{Bhattacharjee_2022_Krylov}. By performing a gradient expansion of $b(n)$, one turns a system \eqref{system} of ordinary  equations to a single partial differential equation on the function $\varphi(n,t)$. Usually, subleading terms of the gradient expansion of $b(n)$ are disregarded \cite{Parker_2019,Barbon_2019_On_evolution,Muck_2022_Krylov,Bhattacharjee_2022_Krylov}. Our findings show that this approximation can be unacceptably rough and miss even the basic features of the autocorrelation function. For example, applying the results of the continuum approximation of ref.~\cite{Bhattacharjee_2022_Krylov} to the exactly solvable model \eqref{bn2}, one gets $C(t)\propto e^{-t}$ at large times. This estimate misses the oscillating prefactor present in the actual large-time approximation $C(t)\simeq 2\cos(\omega t)\, e^{-t}$, see eq.~\eqref{Ali}.

\appendix
\section{Continuous Hahn polynomials \label{appendix}}

The continuous Hahn polynomials	are defined by the relation \cite{askey_continuous_1985,Koekoek_2010}
\begin{align}
p_n(x)=i^n \frac{(a+c)_n (a+d)_n}{n!}\pFfulq{3}{2}{-n,n+a+b+c+d-1,a+i x}{a+c,a+d}{1}\,.
\end{align}
They depend on four parameters, $a$, $b$, $c$, and $d$. We consider the parameters that obey
\begin{align} \label{case}
\mathrm{Re}(a,b,c,d)>0\,,\qquad c=a^*\,,\qquad d=b^*\,.
\end{align}
In this case the orthogonality relation of polynomials is given by the integral over the real axis of the form
\begin{align}
&\frac{1}{2\pi}\int_{-\infty}^{\infty} dx\;\! \Gamma(a+ix)\Gamma(b+ix)\Gamma(c-ix)\Gamma(d-ix)p_m(x)p_n(x)\notag\\
&=\frac{\Gamma(n+a+c)\Gamma(n+a+d) \Gamma(n+b+c)\Gamma(n+b+d)}{(2n+a+b+c+d-1)\Gamma(n+a+b+c+d-1)n!}\delta_{m,n}\,.
\end{align}
Using the Barnes integral
\begin{align}
\frac{1}{2\pi}\int_{-\infty}^{\infty} dx\;\! \Gamma(a+ix)\Gamma(b+ix)\Gamma(c-ix)\Gamma(d-ix)=\frac{\Gamma(a+c) \Gamma(a+d) \Gamma(b+c) \Gamma(b+d)}{\Gamma(a+b+c+d)}\,,
\end{align}
valid at  $\mathrm{Re}(a,b,c,d)>0$, we can introduce the weight as
\begin{align}\label{eq:weightw}
w(x)=\frac{\Gamma(a+b+a^*+b^*)}{2\pi\Gamma(a+a^*) \Gamma(a+b^*) \Gamma(a^*+b) \Gamma(b+b^*)} |\Gamma(a+ix)\Gamma(b+ix)|^2\,,
\end{align}
such that it is normalized, $\int_{-\infty}^{\infty}dxw(x)=1$. The normalization of the polynomials then takes the form
\begin{align}
&\int_{-\infty}^{\infty} dx w(x)p_m(x)p_n(x)\notag\\
&=\frac{(a+c)_n(a+d)_n (b+c)_n(b+d)_n}{(a+b+c+d)_n}\frac{n+a+b+c+d-1}{n!(2n+a+b+c+d-1)}\delta_{m,n}\,.
\end{align}
The recurrence relation is
\begin{align}\label{eq:recurrsion}
p_{n+1}(x)=\frac{B_n}{A_n}[-x+i(A_n+C_n+a)]p_{n}(x) + \frac{B_{n-1}B_n C_n}{A_n}p_{n-1}(x)\,,
\end{align}
where
\begin{gather}
A_n=-\frac{(n+a+b+c+d-1)(n+a+c)(n+a+d)}{(2n+a+b+c+d-1)(2n+a+b+c+d)}\,,\\
B_n=\frac{(n+a+c)(n+a+d)}{n+1}\,,\\
C_n=\frac{n(n+b+c-1)(n+b+d-1)}{(2n+a+b+c+d-2)(2n+a+b+c+d-1)}\,.
\end{gather}

\subsection{Fourier transform}

Consider the Fourier transform involving $p_n(x)$ of the form
\begin{align}\label{eq:Wn}
W_n(t)=\int_{-\infty}^{\infty} dx e^{-ixt}w(x)p_n(x)\,.
\end{align}
Let us consider some general properties of this relation. (i) From the orthogonality of the polynomials it follows $W_n(0)=\delta_{n,0}$. (ii) From the definition it follows $W_{-1}(t)=0$ and $W_0(t)=\int_{-\infty}^{\infty} dx e^{-ixt}w(x)$. Once all $W_j(t)$ are known for $j\le n$, the term $W_{n+1}(t)$ can be found from the recurrence relation (\ref{eq:recurrsion}) and the equality
\begin{align}
\int_{-\infty}^{\infty} dx e^{-ixt}w(x)x p_n(x)=i \frac{\partial W_n(t)}{\partial t}\,.
\end{align}
It yields
\begin{align}\label{eq:Wnrec}
W_{n+1}(t)={}&i\frac{B_n}{A_n}\left[a+A_n+C_n-\frac{\partial}{\partial t} \right]W_n(t)+\frac{B_{n-1}B_n C_n}{A_n}W_{n-1}(t)\,.
\end{align}
Equation (\ref{eq:Wnrec}) is a convenient way to obtain recursively the expressions for $W_{n+1}(t)$ from the preceding two.

\subsection{Transformation to the form of eq.~(\ref{system})}

Let us transform eq.~(\ref{eq:Wnrec}) to the form
\begin{align}\label{eq:maineq}
\frac{\partial \phi_n(t)}{\partial t}=-\tilde b_{n+1} \phi_{n+1}(t)+\tilde a_n\phi_n(t)+\tilde b_n\phi_{n-1}(t)\,.
\end{align}
Introducing
\begin{align}\label{eq:Wphi}
W_n(t)=(-i)^n \alpha_n\phi_n(t)\,,
\end{align}
we obtain the coefficients
\begin{align}
\tilde a_n=a+A_n+C_n\,,\qquad \tilde b_n=B_{n-1}C_n \frac{\alpha_{n-1}}{\alpha_n}\,,
\end{align}
provided
\begin{align}
\left(\frac{\alpha_{n+1}}{\alpha_{n}}\right)^2=-B_{n}^2\frac{C_{n+1}}{A_{n}}
\end{align}
is satisfied for all nonnegative integers $n$. Since $A_n<0$ and $B_n,C_{n+1}>0$ at $n\ge 0$, the solution can be taken as
\begin{align}
\alpha_0=1\,,\qquad \alpha_n=\prod_{j=0}^{n-1} B_j\sqrt\frac{C_{j+1}}{{-A_j}}\,,\qquad n\ge 1\,.
\end{align}
Then we find
\begin{align}
\tilde b_n=\sqrt{-A_{n-1} C_n}\,.
\end{align}

The functions $\phi_n(t)$ can be generated using the recurrence relation
\begin{align}\label{eq:recurrencephi}
\phi_{n+1}(t)=\frac{1}{\tilde b_{n+1}}\left(\tilde a_n-\frac{\partial}{\partial t}\right)\phi_n(t) +\frac{\tilde b_n}{\tilde b_{n+1}}\phi_{n-1}(t)\,,\qquad n\ge0\,.
\end{align}
Here we should use
\begin{align}\label{incnd}
\phi_{-1}(t)=0\,,\qquad \phi_0(t)=\int_{-\infty}^{\infty}dx e^{-i t x}w(x)\,.
\end{align}

\subsection{The case $\tilde a_n=0$}

In the following we consider the case $\tilde a_n=0$, which occurs if $A_n+C_n=-a$. For real $a$, $b$, $c$, and $d$ the admissible set is the one with
\begin{align}\label{case1}
a=c>0\,,\qquad b=d>0\,.
\end{align}
Another possibility with $\tilde a_n=0$ involves the complex parameters and we set $c=a^*$, $d=b^*$. Direct inspection shows that $\tilde a_n=0$ occurs if $\mathrm{Im}(a+b)=0$ and $\mathrm{Re}(a-b) \times \mathrm{Re}(a+b-1)=0$. Therefore two cases arise. One is
\begin{align}\label{case2}
a=d=r+i\omega\,,\qquad b=c=r-i\omega\,,\qquad r>0\,,\qquad \omega\neq 0\,.
\end{align}
The other is
\begin{align}\label{case3}
a=r+i\omega\,,\qquad b=1-r-i\omega\,,\qquad c=r-i\omega\,,\qquad d=1-r+i\omega\,,\qquad 0<r<1\,.
\end{align}

\subsection{Evaluation of $\phi_n(t)$}

Let us evaluate $\phi_n(t)$ for $\tilde a_n=0$. We will use the integral representation of continuous Hahn polynomials \cite{koelink_jacobi_1996}
\begin{align}\label{eq:koelinkidentity}
p_n(z)={}&\frac{i^n}{2^{a+b^*-1}}\frac{\Gamma(a+b^*+n)}{\Gamma(a+iz)\Gamma(b^*-iz)}\notag\\
&\times \int_{-\infty}^{\infty}dx e^{-i 2xz}(1-\tanh x)^a (1+\tanh x)^{b^*} P_n^{(a+a^*-1,b+b^*-1)}(\tanh x)\,,
\end{align}
where  $P_n^{(\alpha,\beta)}(x)$ are the Jacobi polynomials. The latter are defined by
\begin{align}\label{eq:jacobi}
P_n^{(\alpha,\beta)}(x)=\frac{\Gamma(\alpha+n+1)}{n!\;\!\Gamma(\alpha+\beta+n+1)} \sum_{k=0}^{n}(-1)^k \binom{n}{k}\frac{\Gamma(\alpha+\beta+n+k+1)}{\Gamma(\alpha+k+1)}\left(\frac{1-x}{2}\right)^k\,.
\end{align}
We consider $\mathrm{Re}(\alpha,\beta)>-1$, which is fulfilled for $\mathrm{Re}(a,b)>0$ that is assumed in eq.~(\ref{eq:koelinkidentity}). Using
\begin{align}
\int_{-\infty}^{\infty} dz e^{-i xz} \Gamma(b+i z)\Gamma(a^*-i z)=\frac{2\pi\Gamma(a^*+b)}{(1-e^{-x})^b(1+e^x)^{a^*}}\,,
\end{align}
we obtain
\begin{align}
W_n(t)={}&i^n (a+b^*)_n\frac{\Gamma(a+b+a^*+b^*)}{\Gamma(a+a^*) \Gamma(b+b^*)}e^{bt}\notag\\
&\times \int_{0}^{1} dx \frac{(1-x)^{a+a^*-1} x^{b+b^*-1}}{[1-(1-e^{t})x]^{a^*+b}} P_n^{(a+a^*-1,b+b^*-1)}(2x-1)\,.
\end{align}
Using eq.~(\ref{eq:jacobi}) and the integral representation\footnote{\url{https://dlmf.nist.gov/15.6.E1}}
\begin{align}
\pFq{2}{1}{a,b}{c}{z}=\frac{\Gamma(c)}{\Gamma(b)\Gamma(c-b)}\int_0^1 dx \frac{x^{b-1} (1-x)^{c-b-1}}{(1-z x)^a}\,,\qquad \mathrm{Re}(c)>\mathrm{Re}(b)>0\,,
\end{align}
we obtain
\begin{align}\label{eq:Wnjacobi}
W_n={}&\frac{i^n}{n!} (a+b^*)_n (a+a^*)_n e^{bt}\notag\\
&\times\sum_{k=0}^n (-1)^k \binom{n}{k} \frac{(n-1+a+b+a^*+b^*)_k}{(a+b+a^*+b^*)_k} \pFq{2}{1}{a^*+b,b+b^*}{a+b+a^*+b^*+k}{1-e^t}\,.
\end{align}

The initial term $\phi_0(t)$ of eq.~(\ref{incnd}) is identical to $W_0(t)$, see eq.~(\ref{eq:Wn}). Therefore from eq.~(\ref{eq:Wnjacobi}) we obtain
\begin{align}\label{wwqq}
\phi_0(t)=e^{bt}\pFq{2}{1}{a^*+b,b+b^*}{a+b+a^*+b^*}{1-e^t}\,.
\end{align}
Here $\pFq{2}{1}{a,b}{c}{z}$ is the Gauss hypergeometric function defined in eq.~(\ref{eq:2F1}). Equation (\ref{wwqq}) gives complex $\phi_0(t)$ for the case (\ref{case3}). On the other hand, we want to study real $\phi_0(t)$, which occurs in the cases (\ref{case1}) and (\ref{case2}). In the following we thus consider the latter two sets of parameters. Noting that in this case we have
\begin{align}
\alpha_n=\sqrt{\frac{(2a)_n(2b)_n[(a+b)_n]^2 (2a+2b+n-1)_n}{n! (2a+2b)_{2n}}}\,,
\end{align}
from eqs.~(\ref{eq:Wphi}) and (\ref{eq:Wnjacobi}) we obtain
\begin{align}\label{eq:phisum}
\phi_n(t)={}&(-1)^n \sqrt{\frac{(2a+2b)_{2n}(2a)_n}{n!(2b)_n (2a+2b-1+n)_n}}e^{bt}\notag\\
&\times \sum_{k=0}^n (-1)^k \binom{n}{k} \frac{(2a+2b-1+n)_k}{(2a+2b)_k} \pFq{2}{1}{a+b,2b}{2a+2b+k}{1-e^t}\,.
\end{align}
Equation~(\ref{eq:phisum}) contains a finite sum that we were not been able to calculate directly. Instead of that, we found an alternative route to calculate $\phi_n(t)$. The case $n=0$ of eq.~(\ref{eq:phisum}) can be expressed as\footnote{\url{https://functions.wolfram.com/07.23.17.0101.01}}
\begin{align}
\phi_0(t)=\pFq{2}{1}{a,b}{a+b+\frac{1}{2}}{-\sinh^2(t/2)}\,.
\end{align}
Using the recurrence relation (\ref{eq:recurrencephi}),
we then calculated $\phi_1(t)$, $\phi_2(t)$, $\phi_3(t)$, etc., and concluded that they satisfy  the general form
\begin{align}\label{eq:solution1}
\phi_n(t)=\left(\prod_{j=1}^n
\tilde b_j\right) \frac{2^n}{n!} [\sinh(t/2)]^n \pFq{2}{1}{a+n/2,b+n/2}{a+b+\frac{1}{2}+n}{ -\sinh^2(t/2)}\,.
\end{align}
This is our final expression for $\phi_n(t)$.

The properties of the Gauss hypergeometric function enabled us to transform  eq.~(\ref{eq:solution1}) to several other forms. Some of them are given by
\begin{align}
\phi_n(t)=&\left(\prod_{j=1}^n \tilde b_j\right)\frac{2^n}{n!} {[\sinh(t/2)]^n} \pFq{2}{1}{2a+n,2b+n}{a+b+\frac{1}{2}+n}{ -\sinh^2(t/4)}
\end{align}
and
\begin{align}\label{repsumrule}
\phi_n(t)={}&\left(\prod_{j=1}^n \tilde b_j\right)\frac{1}{n!} e^{bt}(e^{t}-1)^n \pFq{2}{1}{a+b+n,2b+n}{2a+2b+2n}{1-e^{t}}\,.
\end{align}
In the main text, two other forms are given. Note that there are several forms of the prefactor that have some similarities with the structure of arguments in the hypergeometric functions. Some forms of the prefactor are
\begin{align}
\prod_{j=1}^n \tilde b_j={}& \left(\frac{n!}{4^{2n}}\frac{(2a)_n(2b)_n(2a+2b-1)_n}{(a+b-1/2)_n(a+b+1/2)_n}\right)^{1/2}\notag\\
={}&\left(\frac{n!(2a)_n (2b)_n (a+b)_n}{2^{2n}(a+b+1/2)_n (2a+2b+n-1)_n}\right)^{1/2}\notag\\
={}&\left(\frac{n!(2a)_n (2b)_n[(a+b)_n]^2}{(2a+2b)_{2n} (2a+2b+n-1)_n}\right)^{1/2}\,.
\end{align}
Using the Clausen identity
\begin{align}
\left[\pFq{2}{1}{a,b}{a+b+\frac{1}{2}}{z}\right]^2 = \pFfulq{3}{2}{2a,2b,a+b}{a+b+\frac{1}{2},2a+2b}{z}\,,
\end{align}
we note another interesting representation
\begin{align}\label{eq:3F2}
\phi_n(t)={}&\left(\frac{(2a)_n (2b)_n (a+b)_n}{n!(a+b+1/2)_n (2a+2b+n-1)_n}\right)^{1/2}  [\sinh(t/2)]^n \notag\\
&\times\left[\pFfulq{3}{2}{2a+n,2b+n,a+b+n}{a+b+n+\frac{1}{2},2a+2b+2n}{ -\sinh^2(t/2)}\right]^{1/2}\,.
\end{align}
The proof that $\phi_n(t)$ indeed satisfies eq.~(\ref{eq:recurrencephi}) is given in the main text.

As a side result of previous considerations, a comparison of eq.~(\ref{eq:phisum}) and the expression (\ref{repsumrule}) leads to the identity
\begin{align}\label{eqsum}
\sum_{k=0}^n (-1)^k \binom{n}{k} \frac{(2a+2b-1+n)_k}{(2a+2b)_k} \pFq{2}{1}{a+b,2b}{2a+2b+k}{z}\notag\\
= \frac{(a+b)_n (2b)_n}{(2a+2b)_{2n}}z^n\pFq{2}{1}{a+b+n,2b+n}{2a+2b+2n}{z}\,.
\end{align}
The sum in the left-hand side of eq.~(\ref{eqsum}) can be found in the literature, see the expression 5.3.5.3 in ref.~\cite{prudnikov3}. It is however expressed in terms of a $_3F_2$ hypergeometric function. After comparing with the right-hand side of eq.~(\ref{eqsum}) we obtain the equality
\begin{align}\label{3f2}
\frac{(a+b)_n (2b)_n}{(2a+2b+n)_{n}}z^n\pFq{2}{1}{a+b+n,2b+n}{2a+2b+2n}{z}= {(1-n)_n}\times\pFfulq{3}{2}{1,a+b,2b}{1-n,2a+2b+n}{z}\,.
\end{align}
The right-hand side of eq.~(\ref{3f2}) is not defined at positive integers $n$, but can be understood as
\begin{align}
{(1-n)_n}\times\pFfulq{3}{2}{1,a+b,2b}{1-n,2a+2b+n}{z}=\sum_{k=n}^{\infty} \frac{(a+b)_k(2b)_k}{(2a+2b+n)_k}\frac{z^k}{(k-n)!}\,.
\end{align}
After shifting the index of summation we then obtain the left-hand side of eq.~(\ref{3f2}). Therefore both derivations of $\phi_n(t)$ are consistent.

Let us verify the sum rule (\ref{sum rule}). Using the representation (\ref{repsumrule}) we can perform the summation using the expression 6.7.2.3 from ref.~\cite{prudnikov3}. Note that the sum rule can also be verified using eq.~(\ref{eq:3F2}) and the expression 6.8.1.31 from ref.~\cite{prudnikov3}.

In the notation of the main text, we have used the rescaled time such that $\varphi_n(t)=\phi_n(4t)$ and $b_n=4\tilde b_n$.

\acknowledgments

OL thanks A. Teretenkov and N. Il'in for useful discussions. Work at Laboratoire de Physique Th\'{e}orique was supported in part by the EUR grant NanoX ANR-17-EURE-0009 in the framework of the``Programme des Investissements d'Avenir''.

\bibliographystyle{JHEP}

\begin{thebibliography}{1000}

\bibitem{Liu_2020_Holographic}
H.~Liu and J.~Sonner, \emph{Holographic systems far from equilibrium: a
  review}, \href{https://doi.org/10.1088/1361-6633/ab4f91}{\emph{Rep. Prog.
  Phys.} {\bfseries 83} (2019) 016001}.

\bibitem{Banuls_2009_Matrix}
M.C.~Ba\~nuls, M.B.~Hastings, F.~Verstraete and J.I.~Cirac, \emph{Matrix
  product states for dynamical simulation of infinite chains},
  \href{https://doi.org/10.1103/PhysRevLett.102.240603}{\emph{Phys. Rev. Lett.}
  {\bfseries 102} (2009) 240603}.

\bibitem{orus2019tensor}
R.~Or{\'u}s, \emph{Tensor networks for complex quantum systems},
  \href{https://doi.org/10.1038/s42254-019-0086-7}{\emph{Nat. Phys. Rev.}
  {\bfseries 1} (2019) 538}.

\bibitem{Parker_2019}
D.E.~Parker, X.~Cao, A.~Avdoshkin, T.~Scaffidi and E.~Altman, \emph{A
  {Universal} {Operator} {Growth} {Hypothesis}},
  \href{https://doi.org/10.1103/PhysRevX.9.041017}{\emph{Phys. Rev. X}
  {\bfseries 9} (2019) 041017}.

\bibitem{Mori_1965_Continued-fraction}
H.~Mori, \emph{A continued-fraction representation of the time-correlation
  functions}, \href{https://doi.org/10.1143/PTP.34.399}{\emph{Prog. Theor.
  Phys.} {\bfseries 34} (1965) 399}.

\bibitem{Dupuis_1967_Moment}
M.~Dupuis, \emph{Moment and continued fraction expansions of time
  autocorrelation functions},
  \href{https://doi.org/10.1143/PTP.37.502}{\emph{Prog. Theor. Phys.}
  {\bfseries 37} (1967) 502}.

\bibitem{viswanath2008recursion}
V.S.~Viswanath and G.~M{\"u}ller, \emph{The Recursion Method: Application to
  Many-Body Dynamics}, vol.~23, Springer, Berlin (2008),
  \href{https://doi.org/10.1007/978-3-540-48651-0}{10.1007/978-3-540-48651-0}.

\bibitem{Muck_2022_Krylov}
W.~M\"{u}ck and Y.~Yang, \emph{Krylov complexity and orthogonal polynomials},
  \href{https://doi.org/10.1016/j.nuclphysb.2022.115948}{\emph{Nucl. Phys. B}
  {\bfseries 984} (2022) 115948}.

\bibitem{Liu_1990_Infinite-temperature}
J.-M.~Liu and G.~M\"uller, \emph{Infinite-temperature dynamics of the
  equivalent-neighbor XYZ model},
  \href{https://doi.org/10.1103/PhysRevA.42.5854}{\emph{Phys. Rev. A}
  {\bfseries 42} (1990) 5854}.

\bibitem{Florencio_1992_Quantum}
J.~Florencio, S.~Sen and Z.~Cai, \emph{Quantum spin dynamics of the transverse Ising model in two dimensions},
  \href{https://doi.org/10.1007/BF00694087}{\emph{J. Low Temp. Phys.}
  {\bfseries 89} (1992) 561}.

\bibitem{Zobov_2006_Second}
V.~Zobov and A.~Lundin, \emph{Second moment of multiple-quantum NMR and a
  time-dependent growth of the number of multispin correlations in solids},
  \href{https://doi.org/10.1134/S1063776106120089}{\emph{J. Exp. Theor. Phys.}
  {\bfseries 103} (2006) 904}.

\bibitem{Elsayed_2014_Signatures}
T.A.~Elsayed, B.~Hess and B.V.~Fine, \emph{Signatures of chaos in time series
  generated by many-spin systems at high temperatures},
  \href{https://doi.org/10.1103/PhysRevE.90.022910}{\emph{Phys. Rev. E}
  {\bfseries 90} (2014) 022910}.

\bibitem{Bouch_2015_Complex}
G.~Bouch, \emph{Complex-time singularity and locality estimates for quantum
  lattice systems}, \href{https://doi.org/10.1063/1.4936209}{\emph{J. Math.
  Phys.} {\bfseries 56} (2015) 123303}.

\bibitem{Cao_2021_Statistical}
X.~Cao, \emph{A statistical mechanism for operator growth},
  \href{https://doi.org/10.1088/1751-8121/abe77c}{\emph{J. Phys. A: Math. Theor.}
  {\bfseries 54} (2021) 144001}.

\bibitem{Noh_2021}
J.D.~Noh, \emph{Operator growth in the transverse-field Ising spin chain with
  integrability-breaking longitudinal field},
  \href{https://doi.org/10.1103/PhysRevE.104.034112}{\emph{Phys. Rev. E}
  {\bfseries 104} (2021) 034112}.

\bibitem{Heveling_2022_Numerically}
R.~Heveling, J.~Wang and J.~Gemmer, \emph{Numerically probing the universal
  operator growth hypothesis},
  \href{https://doi.org/10.1103/PhysRevE.106.014152}{\emph{Phys. Rev. E}
  {\bfseries 106} (2022) 014152}.

\bibitem{Uskov_Lychkovskiy_2024_Quantum}
F.~Uskov and O.~Lychkovskiy, \emph{Quantum dynamics in one and two dimensions
  via the recursion method},
  \href{https://doi.org/10.1103/PhysRevB.109.L140301}{\emph{Phys. Rev. B}
  {\bfseries 109} (2024) L140301}.

\bibitem{De_2024_Stochastic}
A.~De, U.~Borla, X.~Cao and S.~Gazit, \emph{Stochastic sampling of operator
  growth dynamics},
  \href{https://doi.org/10.1103/PhysRevB.110.155135}{\emph{Phys. Rev. B}
  {\bfseries 110} (2024) 155135}.

\bibitem{Loizeau_2025_Quantum}
N.~Loizeau, J.C.~Peacock and D.~Sels, \emph{{Quantum many-body simulations with
  PauliStrings.jl}},
  \href{https://doi.org/10.21468/SciPostPhysCodeb.54}{\emph{SciPost Phys.
  Codebases} (2025) 54}.

\bibitem{Loizeau_2025_Codebase}
N.~Loizeau, J.C.~Peacock and D.~Sels, \emph{{Codebase release 1.5 for
  PauliStrings.jl}},
  \href{https://doi.org/10.21468/SciPostPhysCodeb.54-r1.5}{\emph{SciPost Phys.
  Codebases} (2025) 54}.
  
\bibitem{Teretenkov_2025}
A.~Teretenkov, F.~Uskov and O.~Lychkovskiy, \emph{Pseudomode expansion of
  many-body correlation functions},
  \href{https://doi.org/10.1103/PhysRevB.111.174308}{\emph{Phys. Rev. B}
  {\bfseries 111} (2025) 174308}. 

\bibitem{shirokov2025quench}
I.~Shirokov, V.~Hrushev, F.~Uskov, I.~Dudinets, I.~Ermakov and O.~Lychkovskiy,
  \emph{Quench dynamics via recursion method and dynamical quantum phase
  transitions}, {\emph{arXiv:2503.24362} (2025) }.
  
\bibitem{ermakov2025operator}
I.~Ermakov,
  \emph{Operator growth in many-body systems of higher spins}, {\emph{arXiv:2504.07833} (2025) }.

\bibitem{nandy2024quantum}
P.~Nandy, A.S.~Matsoukas-Roubeas, P.~Mart{\'\i}nez-Azcona, A.~Dymarsky and
  A.~del Campo, \emph{Quantum dynamics in Krylov space: Methods and
  applications}, {\emph{arXiv:2405.09628} (2024) }.

\bibitem{baiguera2025quantum}
S.~Baiguera, V.~Balasubramanian, P.~Caputa, S.~Chapman, J.~Haferkamp, M.~P.~Heller, and N.~Y.~Halpern,
``Quantum complexity in gravity, quantum field theory, and quantum information science,''
\href{https://doi.org/10.48550/arXiv.2503.10753}{arXiv:2503.10753}
  
\bibitem{Caputa_2022_Geometry}
P.~Caputa, J.M.~Magan and D.~Patramanis, \emph{Geometry of Krylov complexity},
  \href{https://doi.org/10.1103/PhysRevResearch.4.013041}{\emph{Phys. Rev.
  Research} {\bfseries 4} (2022) 013041}.

\bibitem{Kim_2022_Operator}
J.~Kim, J.~Murugan, J.~Olle and D.~Rosa, \emph{Operator delocalization in
  quantum networks},
  \href{https://doi.org/10.1103/PhysRevA.105.L010201}{\emph{Phys. Rev. A}
  {\bfseries 105} (2022) L010201}.

\bibitem{Adhikari_2022_Cosmological}
K.~Adhikari and S.~Choudhury, \emph{Cosmological Krylov complexity},
  \href{https://doi.org/10.1002/prop.202200126}{\emph{Fortschr. Phys.}
  {\bfseries 70} (2022) 2200126}.

\bibitem{Jian_2021_Complexity}
S.-K.~Jian, B.~Swingle and Z.-Y.~Xian, \emph{Complexity growth of operators in
  the {SYK} model and in {JT} gravity},
  \href{https://doi.org/10.1007/JHEP03(2021)014}{\emph{J. High Energy Phys.}
  {\bfseries 2021} (2021) 1}.

\bibitem{Kar_2022_Random}
A.~Kar, L.~Lamprou, M.~Rozali et~al., \emph{Random matrix theory for complexity
  growth and black hole interiors},
  \href{https://doi.org/10.1007/JHEP01(2022)016}{\emph{J. High Energy Phys.}
  {\bfseries 2022} (2022) 16}.

\bibitem{Adhikari_2023_Krylov}
K.~Adhikari, S.~Choudhury and A.~Roy, \emph{Krylov complexity in quantum field
  theory}, \href{https://doi.org/10.1016/j.nuclphysb.2023.116263}{\emph{Nucl.
  Phys. B} {\bfseries 993} (2023) 116263}.

\bibitem{Avdoshkin_2024_Krylov}
A.~Avdoshkin, A.~Dymarsky and M.~Smolkin, \emph{Krylov complexity in quantum
  field theory, and beyond},
  \href{https://doi.org/10.1007/JHEP06(2024)066}{\emph{J. High Energy Phys.}
  {\bfseries 2024} (2024) 66}.

\bibitem{chapman_krylov_2025}
S.~Chapman, S.~Demulder, D.A.~Galante, S.U.~Sheorey and O.~Shoval, \emph{Krylov
  complexity and chaos in deformed {Sachdev}-{Ye}-{Kitaev} models},
  \href{https://doi.org/10.1103/PhysRevB.111.035141}{\emph{Phys. Rev. B}
  {\bfseries 111} (2025) 035141}.

\bibitem{caputa2024krylov}
P.~Caputa, H.-S.~Jeong, S.~Liu, J.~F.~Pedraza, and L.-C.~Qu,
``Krylov complexity of density matrix operators,''
\href{https://doi.org/10.1007/JHEP05(2024)337}{\emph{J. High Energy Phys.}, \textbf{2024}, 05, 337, 2024}.

\bibitem{camargo2024spectral}
  H.~A. Camargo, V.~Jahnke, H.-S. Jeong, K.-Y. Kim, and M.~Nishida,
  ``Spectral and Krylov complexity in billiard systems,''
    \href{https://doi.org/10.1103/PhysRevD.109.046017}{\emph{Phys. Rev. D}, \textbf{109}, 4, 046017, 2024}.


\bibitem{Dymarsky_2021_Krylov}
A.~Dymarsky and M.~Smolkin, \emph{Krylov complexity in conformal field theory},
  \href{https://doi.org/10.1103/PhysRevD.104.L081702}{\emph{Phys. Rev. D}
  {\bfseries 104} (2021) L081702}.

\bibitem{Caputa_2021_Operator}
P.~Caputa and S.~Datta, \emph{Operator growth in 2d {CFT}},
  \href{https://doi.org/10.1007/JHEP12(2021)188}{\emph{J. High Energy Phys.}
  {\bfseries 2021} (2021) 188}.

\bibitem{dodelson2025black}
M.~Dodelson, \emph{Black holes from chaos}, {\emph{arXiv:2501.06170} (2025) }.


\bibitem{Banchi_2013}
L.~Banchi and R.~Vaia, \emph{Spectral problem for quasi-uniform
  nearest-neighbor chains}, \href{https://doi.org/10.1063/1.4797477}{\emph{J.
  Math. Phys.} {\bfseries 54} (2013) 043501}.

\bibitem{Gamayun_2020}
O.~Gamayun, O.~Lychkovskiy and J.-S.~Caux, \emph{{Fredholm determinants, full
  counting statistics and Loschmidt echo for domain wall profiles in
  one-dimensional free fermionic chains}},
  \href{https://doi.org/10.21468/SciPostPhys.8.3.036}{\emph{SciPost Phys.}
  {\bfseries 8} (2020) 036}.

\bibitem{Ljubotina_2019}
M.~Ljubotina, S.~Sotiriadis and T.~Prosen, \emph{{Non-equilibrium quantum
  transport in presence of a defect: the non-interacting case}},
  \href{https://doi.org/10.21468/SciPostPhys.6.1.004}{\emph{SciPost Phys.}
  {\bfseries 6} (2019) 004}.

\bibitem{Viswanath_1994_Ordering}
V.S.~Viswanath, S.~Zhang, J.~Stolze and G.~M\"uller, \emph{Ordering and
  fluctuations in the ground state of the one-dimensional and two-dimensional
  s=1/2 XXZ antiferromagnets: A study of dynamical properties based on the
  recursion method},
  \href{https://doi.org/10.1103/PhysRevB.49.9702}{\emph{Phys. Rev. B}
  {\bfseries 49} (1994) 9702}.

\bibitem{Yates_2020_Lifetime}
D.J.~Yates, A.G.~Abanov and A.~Mitra, \emph{Lifetime of almost strong edge-mode
  operators in one-dimensional, interacting, symmetry protected topological
  phases}, \href{https://doi.org/10.1103/PhysRevLett.124.206803}{\emph{Phys.
  Rev. Lett.} {\bfseries 124} (2020) 206803}.

\bibitem{Yates_2020_Dynamics}
D.J.~Yates, A.G.~Abanov and A.~Mitra, \emph{Dynamics of almost strong edge
  modes in spin chains away from integrability},
  \href{https://doi.org/10.1103/PhysRevB.102.195419}{\emph{Phys. Rev. B}
  {\bfseries 102} (2020) 195419}.

\bibitem{Bhattacharjee_2022_Krylov}
B.~Bhattacharjee, X.~Cao, P.~Nandy and T.~Pathak, \emph{Krylov complexity in
  saddle-dominated scrambling},
  \href{https://doi.org/10.1007/JHEP05(2022)174}{\emph{J. High Energy Phys.}
  {\bfseries 2022} (2022) 1}.

\bibitem{Camargo_2023_Krylov}
H.A.~Camargo, V.~Jahnke, K.-Y.~Kim and M.~Nishida, \emph{Krylov complexity in
  free and interacting scalar field theories with bounded power spectrum},
  \href{https://doi.org/10.1007/JHEP05(2023)226}{\emph{J. High Energy Phys.}
  {\bfseries 2023} (2023) 1}.
  
\bibitem{lunt2025emergent}
O.~Lunt, T.~Kriecherbauer, K.~T.-R. McLaughlin, and C.~von Keyserlingk,
``Emergent random matrix universality in quantum operator dynamics,''
\href{https://doi.org/10.48550/arXiv.2504.18311}{arXiv:2504.18311v2}

\bibitem{pinna2025approximation}
G.~Pinna, O.~Lunt, and C.~von Keyserlingk,
``Approximation theory for Green’s functions via the Lanczos algorithm,''
\href{https://doi.org/10.48550/arXiv.2505.00089}{arXiv:2505.00089v1}
  
\bibitem{sasaki2024towards}
  R.~Sasaki,
  ``Towards verifications of Krylov complexity,''
    \href{https://doi.org/10.1093/ptep/ptae073}{\emph{Prog. Theor. Exp. Phys.}, \textbf{2024}, 063A01}.

\bibitem{chihara}
T.S.~Chihara, \emph{An Introduction to Orthogonal Polynomials}, Gordon and
  Breach, New York (1978).

\bibitem{TalEzer_1984_Accurate}
H.~Tal-Ezer and R.~Kosloff, \emph{An accurate and efficient scheme for
  propagating the time dependent Schr\"{o}dinger equation},
  \href{https://doi.org/10.1063/1.448136}{\emph{J. Chem. Phys.} {\bfseries 81}
  (1984) 3967}.

\bibitem{Vijay1999}
A.~Vijay, R.E.~Wyatt and G.D.~Billing, \emph{Time propagation and spectral
  filters in quantum dynamics: A Hermite polynomial perspective},
  \href{https://doi.org/10.1063/1.480483}{\emph{J. Chem. Phys.} {\bfseries 111}
  (1999) 10794}.

\bibitem{Chen_1999_Chebyshev}
R.~Chen and H.~Guo, \emph{The Chebyshev propagator for quantum systems},
  \href{https://doi.org/https://doi.org/10.1016/S0010-4655(98)00179-9}{\emph{Comput.
  Phys. Commun.} {\bfseries 119} (1999) 19}.

\bibitem{Weibe_2006_Kernel}
A.~Wei\ss{}e, G.~Wellein, A.~Alvermann and H.~Fehske, \emph{The kernel
  polynomial method},
  \href{https://doi.org/10.1103/RevModPhys.78.275}{\emph{Rev. Mod. Phys.}
  {\bfseries 78} (2006) 275}.

\bibitem{Soares_2024_Non-unitary}
R.D.~Soares and M.~Schir\`{o}, \emph{{Non-unitary quantum many-body dynamics using
  the Faber polynomial method}},
  \href{https://doi.org/10.21468/SciPostPhys.17.5.128}{\emph{SciPost Phys.}
  {\bfseries 17} (2024) 128}.


  
\bibitem{fan2023generalised}
  Z.~Y. Fan,
  ``Generalised Krylov complexity,''
  \emph{arXiv preprint arXiv:2306.16118}, 2023.

\bibitem{Barbon_2019_On_evolution}
J.~Barb\'{o}n, E.~Rabinovici, R.~Shir and E.~Trivella, \emph{On the evolution
	of operator complexity beyond scrambling},
\href{https://doi.org/10.1007/JHEP10(2019)264}{\emph{J. High Energy Phys.}
	{\bfseries 2019} (2019) 264}.

\bibitem{Koekoek_2010}
R.~Koekoek, P.A.~Lesky and R.F.~Swarttouw, \emph{Hypergeometric Orthogonal
  Polynomials and Their q-Analogues}, Springer Berlin Heidelberg (2010),
  \href{https://doi.org/10.1007/978-3-642-05014-5}{10.1007/978-3-642-05014-5}.

\bibitem{Bhattacharjee_2023_Operator}
B.~Bhattacharjee, X.~Cao, P.~Nandy and T.~Pathak, \emph{{Operator growth in
  open quantum systems: lessons from the dissipative SYK}},
  \href{https://doi.org/10.1007/JHEP03(2023)054}{\emph{J. High Energy Phys.}
  {\bfseries 2023} (2023) 1}.

\bibitem{Clarkson2023}
P.A.~Clarkson, K.~Jordaan and A.~Loureiro, \emph{Generalized higher-order Freud
  weights}, \href{https://doi.org/10.1098/rspa.2022.0788}{\emph{Proc. R. Soc.
  A} {\bfseries 479} (2023) 20220788}.

\bibitem{dets}
K.~Dilcher and L.~Jiu, \emph{Orthogonal polynomials and {Hankel} determinants
  for certain {Bernoulli} and {Euler} polynomials},
  \href{https://doi.org/10.1016/j.jmaa.2020.124855}{\emph{J. Math. Anal. Appl.}
  {\bfseries 497} (2021) 124855}.



\bibitem{Wang_2024_Diffusion}
J.~Wang, M.H.~Lamann, R.~Steinigeweg and J.~Gemmer, \emph{Diffusion constants
  from the recursion method},
  \href{https://doi.org/10.1103/PhysRevB.110.104413}{\emph{Phys. Rev. B}
  {\bfseries 110} (2024) 104413}.

\bibitem{askey_continuous_1985}
R.~Askey, \emph{Continuous {Hahn} polynomials},
  \href{https://doi.org/10.1088/0305-4470/18/16/004}{\emph{J. Phys. A: Math.
  Gen.} {\bfseries 18} (1985) L1017}.

\bibitem{koelink_jacobi_1996}
H.T.~Koelink, \emph{On {Jacobi} and continuous {Hahn} polynomials},
  \href{https://doi.org/10.1090/S0002-9939-96-03190-5}{\emph{Proc. Amer. Math.
  Soc.} {\bfseries 124} (1996) 887}.

\bibitem{prudnikov3}
A.P.~Prudnikov, {\BIBYu}.A.~Brychkov and O.I.~Marichev, \emph{Integrals and
  Series, Volume 3}, Gordon and Breach, New York (1990).

\end{thebibliography}

\providecommand{\BIBYu}{Yu}

\providecommand{\href}[2]{#2}\begingroup\raggedright
\endgroup

\end{document}